\documentclass[a4paper, aps, amsfonts, amssymb, amsmath, reprint, showkeys, nofootinbib, twoside, superscriptaddress,prx,floatfix]{revtex4-2}
\usepackage[english]{babel}
\usepackage[utf8]{inputenc}
\usepackage{float}
\usepackage[colorinlistoftodos, color=green!40, prependcaption]{todonotes}
\usepackage{amsthm}
\usepackage{mathtools}
\usepackage{physics}
\usepackage{xcolor}
\usepackage{graphicx}
\usepackage[left=23mm,right=13mm,top=35mm,columnsep=15pt]{geometry} 
\usepackage{adjustbox}
\usepackage{placeins}
\usepackage[T1]{fontenc}
\usepackage{lipsum}
\usepackage{csquotes}
\usepackage[pdftex, pdftitle={Article}, pdfauthor={Author}]{hyperref}
\usepackage{cleveref}
\usepackage{graphicx}
\usepackage[toc,page]{appendix}
\usepackage{placeins}
\usepackage{footmisc} 
\usepackage{soul}

\def\Z{\mathbb{Z}}
\def\Vs{V_q}

\def\u{{\bf u}}
\def\SqVs{\sqrt{\Vs}}
\def\taudiv{\tau_{\rm div}}

\def\el{rep}

\begin{document}
\title{The role of the cell cycle in collective cell dynamics}

\author{Jintao Li}
\affiliation{Department of Chemical Engineering, Kyoto University, Kyoto 615-8510, Japan}

\author{Simon K. Schnyder}
\affiliation{Department of Chemical Engineering, Kyoto University, Kyoto 615-8510, Japan}
\author{Matthew S. Turner}
\email{m.s.turner@warwick.ac.uk}
\affiliation{Department of Physics, University of Warwick, Coventry CV4 7AL, UK}
\affiliation{Department of Chemical Engineering, Kyoto University, Kyoto 615-8510, Japan}
\author{Ryoichi Yamamoto}
\affiliation{Department of Chemical Engineering, Kyoto University, Kyoto 615-8510, Japan}

\date{\today} 

\begin{abstract}
Cells coexist together in colonies or as tissues. Their behaviour is controlled by an interplay between intercellular forces and biochemical regulation. We develop a simple model of the cell cycle, the fundamental regulatory network controlling growth and division, and couple this to the physical forces arising within the cell collective. We analyse this model using both particle-based computer simulations and a continuum theory. We focus on 2D colonies confined in a channel. These develop moving growth fronts of dividing cells with quiescent cells in the interior. The profile and speed of these fronts are non-trivially related to the substrate friction and the cell cycle parameters, providing a possible approach to measure such parameters in experiments. 
\end{abstract}

\maketitle

\section{Introduction} \label{sec:intro}

Cell growth and division underlie embryonic morphogenesis, wound healing, and tumour development  \cite{friedl2009collective,sadati2013collective,garcia2015physics,camley2017physical,lin2017collective,ladoux2017mechanobiology,stuelten2018cell,hamby2019connecting,spatarelu2019biomechanics} while the growth and division of undifferentiated cells controls the large-scale movement of cell colonies  \cite{volfson2008biomechanical,bove2017local,puliafito2012collective,doostmohammadi2015celebrating,heinrich2020size}. The expansion of these cell collectives is also under mechanical control  \cite{trepat2009physical,luo2013molecular,mao2015tug,persat2015mechanical,delarue2016self,wang2017review,irvine2017mechanical,chen2019tensile} in which the fates of individual cells determine the collective dynamics of the whole colony. Nonuniformity of mechanical stresses, nutrient levels, waste byproduct and signalling molecules, further influence growth  \cite{hamouche2017bacillus,yabunaka2017cell,malmi2018cell,srinivasan2019multiphase,ben2019physics}. Growing cell colonies typically display a growth front in advance of a crowded interior that contains nearly immobile cells, as seen in bacteria  \cite{mather2010streaming,dell2018growing}, budding yeast  \cite{marinkovic2019microfluidic} and animal cells  \cite{trepat2009physical,puliafito2012collective,xi2017emergent,huergo2011dynamics,Gauquelin2019,heinrich2020size}. There is also interest in the role of nutrient concentration and metabolic regulation \cite{wang2009metabolism,marinkovic2019microfluidic} although this is not the focus of the present work. 

\begin{figure}
	\centering
	\includegraphics[width=86mm]{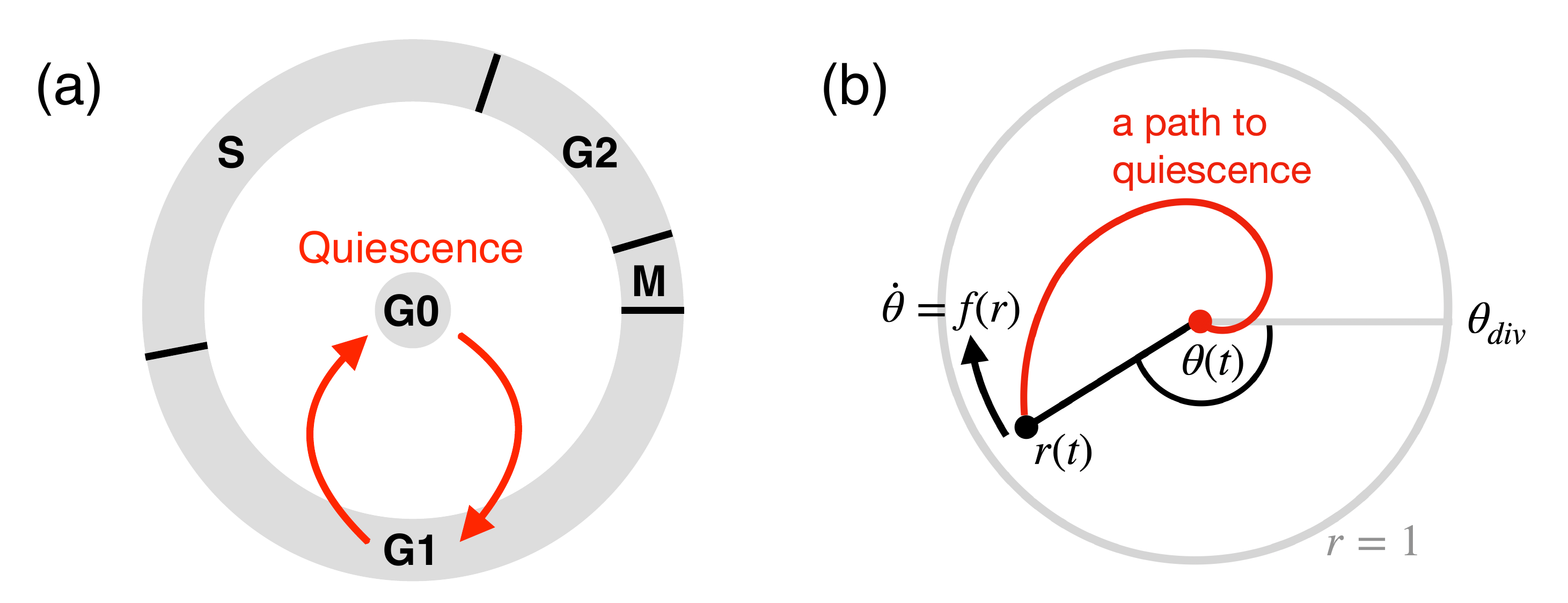}
	\caption{Two representations of the cell cycle. Panel (a) shows a typical biologist's cartoon of the cell cycle, with sequential phases M = Mitosis, G1 = Gap (or Growth) 1, S = Synthesis (of DNA), G2 = Gap 2 and G0 = Gap 0 (Quiescence), a resting state, off the main cycle, in which cells do not divide. Cells undergo mitosis and divide at the end of M phase. The proliferation-quiescence switch is controlled by mechanical stress and other microenvironmental factors. (b): A representation of the cell cycle model studied here. The phase of the cell cycle is parameterised by an angle $\theta$ with a parameter $r$ a proxy for cell cycle activity, represented by a radial distance. The dynamics of $r$ are controlled by a separate equation that is sensitive to physical stresses (see text and \cref{rdotdimensional}). Unstressed, the cell cycle orbits around the circle $r=1$ in a space that can be thought of as a proxy for the biochemical/genetic concentrations, dividing every $\taudiv$ when $\theta=\theta_{\rm div}=2n\pi$ with $n\in \Z$, just like in (a). The red line represents a (reversible) transition between a proliferating state and full quiescence at $r=0$ with the rate of progress around the cycle $f(r)$ a decreasing function of activity $r$, reflecting slowing of the cell cycle. }
	\label{fig:cellCycle}
\end{figure}

Cell proliferation is controlled by the cell cycle, a genetic regulatory mechanism. To divide, cells must pass so-called ``checkpoints'' to ensure that the mother cell is properly prepared to divide. Fig.~\ref{fig:cellCycle} (a) shows a schematic of the cell cycle, which is classified into four phases, each characterized by a set of separate events: growth and preparation for DNA replication (G1), DNA replication (S), preparation for mitosis (G2), and mitosis (M), immediately followed by cell division. There are three main checkpoints: The G1 checkpoint control mechanism ensures that the cell is ready for DNA synthesis; the G2 checkpoint ensures that the cell is ready to enter the M (mitosis) phase and divide; a checkpoint in the middle of mitosis (Metaphase Checkpoint) ensures that the cell is ready to complete cell division. Cells that have temporarily stopped dividing are said to have entered a quiescence state, also called the G0 phase. The switch into this state usually happens in the G1 phase and is reversible, should conditions again become more favourable for growth and division. Cells that are densely packed, small and physically or chemically stressed tend to divide more slowly, corresponding to a slowing of the underlying cell cycle, sometimes to full quiescence  \cite{theveneau2013collective,benham2015mechanical}.

The cell cycle involves an interplay between a range of genetic promoters and inhibitors, with analogues found across most eucaryotes. For example, upon receiving a pro-mitotic extracellular signal, G1 cyclins and cyclin-dependent kinases (CDKs) become active in preparing the cell for S phase, promoting the expression of transcription factors and enzymes required for DNA replication. In contrast, two families of genes, the cip/kip (CDK interacting protein/Kinase inhibitory protein) family and the INK4a/ARF (Inhibitor of Kinase 4/Alternative Reading Frame) family, prevent the progression of the cell cycle by binding to and inactivating the corresponding cell-cycle promoters  \cite{morgan2007cell}. Quiescent cells maintain a transcriptional state that is  different from proliferating cells, achieved by restraining proliferation and cell-cycle progression genes  \cite{fiore2018sleeping}. A growing body of evidence suggests that quiescence is a non-terminal and tissue-specific state that can be initiated and sustained by mechanical factors, such as cell-to-cell friction and extracellular matrix (ECM) friction. {Cells can sense external physical cues. Physical forces are transmitted via biochemical signaling pathways that regulate the cell cycle\cite{mathieu2019intracellular,uroz2018regulation, chen2019tensile}.} On the contrary, some recent experiments examining cellular protein dynamics have shown that processes such as protein expression and transcription are related to the cell cycle state. In other words,  the cell cycle is not only a consequence of cell dynamics, but the cell cycle state itself influences cell behavior\cite{otto2017cell,kent2019broken,Devany2021}.

Underlying this complexity is a simple picture: the cell cycle can be viewed as a biochemical oscillator, producing almost regular division events if the cell is not otherwise perturbed, e.g. by lack of nutrient or physical stresses. Similar oscillators are known to produce circadian rhythms in plants \cite{LOCKE2005383,creux2019circadian} and there are also analogues in abiotic chemical systems {\it in vitro}, such as the well-known BZ reaction  \cite{engquist2016encyclopedia}. Such chemical oscillators require a minimum of two components, each influencing the production of the other, although {\it in vivo} there are many more than this. The course of these reaction can be visualised as a trajectory in the space of the concentrations of the various chemical species (on each axis). These trajectories form closed loops, signifying the presence of an oscillator\footnote{In closed chemical systems a better description would be that of a spiral, as the reactive components are gradually depleted.}. A typical cartoon of the cell cycle, as shown in Fig~\ref{fig:cellCycle}, can also be seen as a simplified projection of these closed reaction loops. The network that controls the same process in prokaryotes is similar in function but differs in its molecular components and checkpoints, e.g. no nuclear envelope breakdown (or reformation) is necessary.

Cell biologists have been studying the cell cycle for many decades  \cite{baserga1968biochemistry,schafer1998cell,vermeulen2003cell, golias2004cell,alberts2015essential,matson2017cell}. Historically, these studies sought to minimise the effect of mechanical heterogeneity and study the cell cycle ``in isolation'' as far as that is possible. This philosophy is now rapidly changing, with a direct role for mechanics experimentally confirmed  \cite{li2014landscape,benham2015mechanical,gao2017cell,takao2019mechanical}. Several key signalling pathways have been identified to correlate with mechanical-feedback capacity  \cite{wagstaff2016mechanical,mascharak2017yap,raj2019reciprocal,perez2019yap}. This is driving interest in the role of mechanical signalling, e.g. in cellular morphology, colony development and cell competition  \cite{vining2017mechanical,chu2018self,saeed2019finite,levayer2020solid}.  Alongside this there has been increasing interest in active tissues in the physics community, with the development of models of active, out of equilibrium materials that can be reminiscent of foams or soft glassy materials \cite{alt2017vertex,malmi2018cell,fiore2018sleeping,srinivasan2019multiphase,alert2020physical} and commonly employ continuum hydrodynamic analysis  \cite{Blanch-Mercader2017,Ishihara2017,yabunaka2017cell,julicher2018hydrodynamic,williamson2018stability, banerjee2019continuum,xi2019material,alert2020physical}. Understanding the physical behaviour of the tissue is the main motivating factor and cell division is typically included in a fairly stylised form \cite{schnyder2020control}, if at all. The state of the art in this respect  \cite{malmi2018cell} incorporates cell elasticity and adhesive cell-cell interactions, as well as cell birth and death but can't really be said to incorporate any description of the cell cycle.

\subsection{Motivation and outline}

Our goal in the present work is a model that couples the physical and biochemical/genetic descriptions: the physics affects the biochemistry in the way mechanical stress slows the cell cycle, while the cell cycle affects the mechanical stresses because it controls cell proliferation and volume, in turn driving the flows that determine these stresses. The motivation for the present work is to combine a functional, rather than biochemically specific, model of the cell cycle with a realistic physical model. We aim to keep the structure of this model as simple as possible. In particular we assume (i) constant, uniform nutrient levels,  (ii) waste or signalling compounds are rapidly removed/diluted and (iii) the role of both motility/migration and apoptosis (cell death) are initially neglected. Our motivation in neglecting these factors is to isolate the role of the cell cycle as far as possible in this study so that it can be studied in relative isolation. In section \ref{sec:discussion} we outline how these other factors could be incorporated within the same framework to generalise our model.

We propose a minimal model that involves a cell cycle oscillator coupled to a physical model through the local stress (pressure) and cell volume. Our  primary goal  is to show that the cell cycle, and the parameters that underlie it, can non-trivially influence the dynamics of growing colonies, as well as the distribution of cell cycle activity, cell volume and stress within it. We propose the basic governing equations in section \ref{sec:model} which can be inferred to be reasonable starting points for a stylised model that reflect previous experimental studies: cell cycle progression control\cite{novak2007irreversible,lopez2009irreversibility}, mechanical-feedback regulation of cell proliferation state\cite{streichan2014spatial,duszyc2018mechanosensing,mathieu2019intracellular,rizzuti2020mechanical}, and cell size regulation during cell division or quiescence\cite{puliafito2012collective, ginzberg2018cell, vargas2018cell}. For simplicity we focus on colony expansion in a simple quasi 2D channel geometry and provide a mechanism for relating the speed and structure of the growth front to the parameters controlling the cell cycle. We first encode our model in a particle-based simulation in which each cell carries its own cell-cycle oscillator controlling size and division events. We then compare this with a continuum analysis that offers substantial analytic insight and a number of scaling results, including an inverse square-root scaling of the front speed with the substrate friction. We obtain quantitative to semi-quantitative agreement between this continuum analysis and our simulations, validating the continuum model. We then compare our model with recent experiments on MDCK cells and find encouraging agreement with the cell area distributions and division rates, in spite of the fact that motility likely plays a role in the motion of these cells, except possibly at the lowest surface frictions.

We simulate colony expansion using a range of parameters to show how the speed and structure of the front depends on the cell cycle variables and show how this can be inverted to extract biophysical parameters from experimental measurements. 

\section{Model}\label{sec:model}
We propose a mechanical-feedback cell cycle model, which in which cells can reversibly switch between proliferation and quiescence driven by local pressure. Here the progress around the cell cycle is measured by an angular variable $\theta$, with division at  $\theta=2n\pi$ with $n\in\Z$, obeying
\begin{equation}\frac{\partial\theta}{\partial t}=f(r)\label{thetadotdimensional}\end{equation}
with $r$ a proxy for the cell cycle activity and take $f(r)=\omega_0 r$ for simplicity. 
An unstressed cell at $r=1$ divides with a division time $\taudiv = 2\pi/\omega_0$.
The cell cycle activity $r$ is assumed to depend on the local pressure $p$ according to
\begin{equation}\frac1r \frac{\partial r}{\partial t}=g(r,p)\label{rdotdimensional}\end{equation}
and is therefore suppressed when $p>0$, eventually approaching $r=0$ if $p\ge p_r$.
For simplicity we take $g(r,p)=\frac1{\tau_r}\left(1-r-p/p_r\right)$ with $\tau_r$ a characteristic time on which the cell cycle responds to mechanical perturbation and where sensitivity is controlled by a reference pressure $p_r$. There is a fixed point at $r=1$ for vanishing pressure $p=0$, corresponding to an unstressed cell cycle that undergoes circular orbits at $r=1$ in a space that can be thought of as a proxy for the biochemical/genetic concentrations. 

Except at division events the volume of any cell is assumed to be given by
\begin{equation}\frac1V \frac{\partial V}{\partial t}= c \frac{\partial\theta}{\partial t}-\frac1{\tau_v}(V-\Vs)/\Vs\label{Vdotdimensional}\end{equation}
This encodes cellular growth with a rate assumed proportional to its rate of progress around the cell cycle $\frac{\partial\theta}{\partial t}$ via a dimensionless growth rate $c$, but also that the cell volume reverts to a minimal quiescent volume $\Vs$ if its cell cycle stalls, with a characteristic recovery time given by $\tau_v$.

We nondimensionalise according to $\tilde t=\omega_0t/(2\pi)$, $\tilde\tau_i=\omega_0\tau_i/(2\pi)$, 
$\tilde c=2\pi c /\omega_0$,
$\tilde p=p/p_r$, 
and $\tilde V=V/\Vs$, hence all lengths are measured in units of $\sqrt{\Vs}$ in 2D. Using dot $(\dot{})$ to indicate $\frac\partial{\partial\tilde t}$ and then dropping all $\tilde{}$ for convenience we have 
\begin{equation}\dot\theta=2\pi r\label{thetadot}\end{equation}
\begin{equation}\tau_r\frac{\dot r}r=1-r-p\label{rdot}\end{equation}
\begin{equation}\frac{\dot V}V= c r-\frac1{\tau_v}(V-1)\label{Vdot}\end{equation}
Eq~\ref{Vdot} only applies between division events and so we use the symbol $v$ to denote the average volume, in the presence of division, as measured in our simulations and for consistency with later continuum analysis. In these dimensionless units the cell cycle is parameterised by only three variables: a growth rate $c$ and recovery times for the cell cycle and volume, $\tau_r$ and $\tau_v$ respectively; the reference pressure $p_r$ will also prove important in providing a reference scale for hydrodynamic/frictional stresses, arising from flows. 
It's worth noting that $\tau_r$ and $\tau_v$ can be experimentally estimated by measuring the time length spend for a cell subjected to growth-limiting pressure to  (i) stop dividing or (ii) finish shrinking to the quiescent volume\cite{puliafito2012collective,heinrich2020size}. In addition to the phenomenological description, we found the timescales can also be linked to molecular processes. For example, the way $\tau_r$ leads to a lag in division response resembles the p53 tumor suppressor's activation can lead to later cell cycle arrest\cite{engeland2018cell}, while $\tau_r$ reflects the period of inhibition of YAP/TAZ  preventing uncontrolled cell volume growth\cite{perez2018cell}.

\subsection{Simulation methods}\label{simulation-methods}

\begin{figure}
	\centering
	\includegraphics[width=86mm]{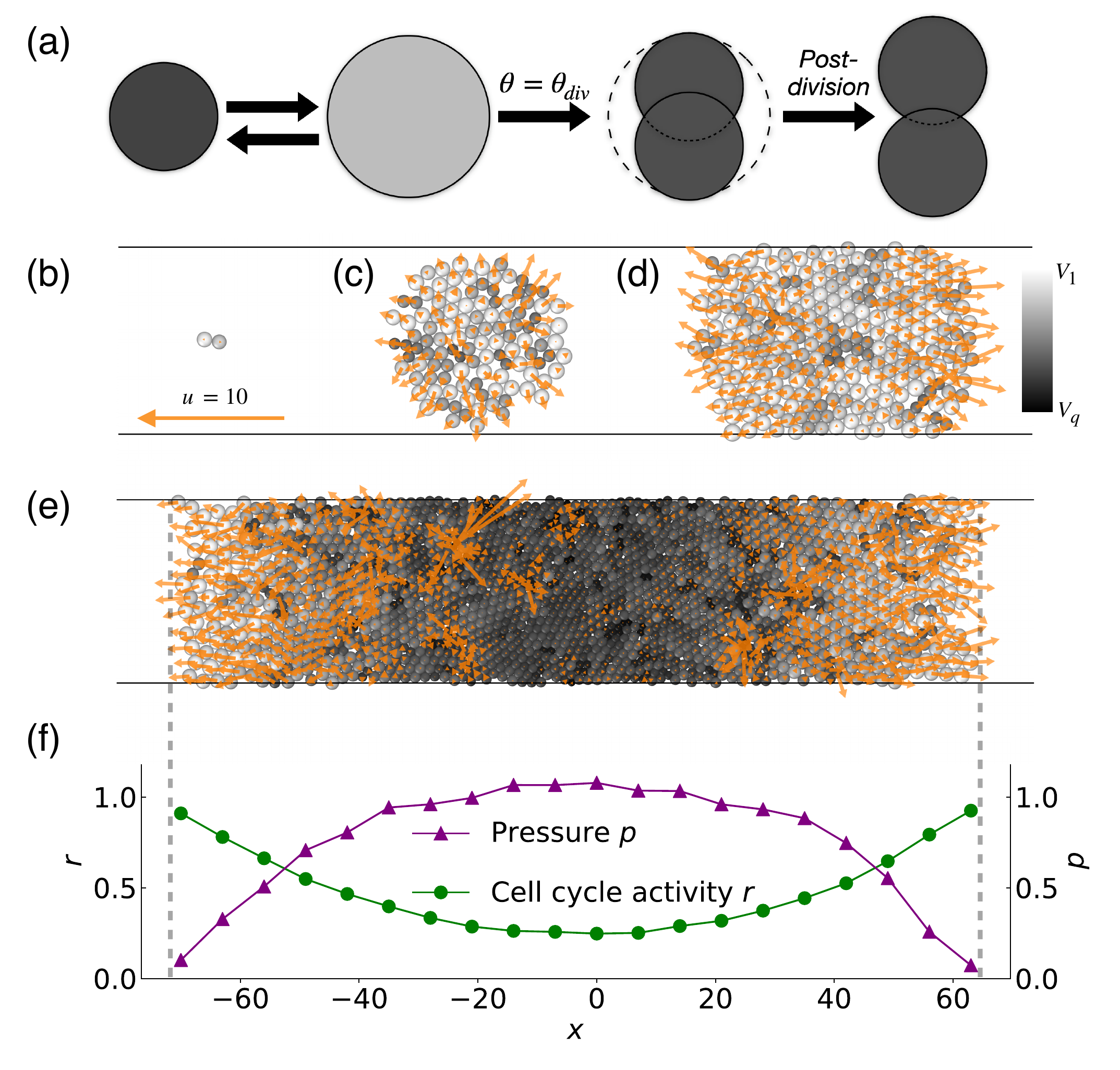}
	\caption{Overview of 2D particle-based simulations. (a) Cartoon of cell division and growth, with darker colours indicating smaller volumes throughout. Cells grow in size when the cell cycle activity is high but can shrink if the cell cycle slows, due to elevated local pressure according to \cref{Vdot} (left/right arrows). When (if) the cell cycle reaches M phase at $\theta_{div}=2n\pi$ the cell divides. Each daughter cell inherits half the mother's current volume (area in 2D), initially overlapping so as to occupy the mother's spatial footprint. The daughters repel one-another and rapidly separate. (b)-(d) The colony grows in time, from just after the first cell divides in (b) reaching approximately 1500 cells in (d). Yellow arrows represent instantaneous cell velocities;  velocity $u=10$ (in dimensionless units of $\sqrt{\Vs}/\taudiv$) shown in key on panel (b). (e) Growth fronts later develop at the left/right edges of the colony while the interior bulk contains smaller cells that rarely divide. (f) The instantaneous distribution of dimensionless pressure $p$ and cell cycle activity $r$ along the channel as shown in (e); in the bulk the pressure is falling towards the cell cycle reference (stalling) pressure driving the cell cycle activity down according to \cref{rdot}, as the cells approach quiescence. The grey dashed lines connecting (e) and (f) mark the colony edges. In these, and all later figures unless noted otherwise, we employ a dimensionless surface friction $\zeta=0.01$, $\tau_r=\tau_v=\taudiv$, and a growth rate corresponding to a nominal average cell volume $v_1=5\Vs$ five times the quiescent volume (see appendix for further details).}
	\label{fig:method}
\end{figure}

We employ a particle-based simulation in which each cell, with unique index $i$, carries its own local variables $\theta_i(t)$, $r_i(t)$ and $v_i(t)$ with it. On division the mother cell volume is divided equally between two daughters that inherit their mother's cell cycle phase $\theta$ and activity $r$, see Fig~\ref{fig:method}a. 
As we will be focusing on 2D colony growth in the present work, cells are treated as elastic disks with a time-dependent natural radius to accommodate growth (and division), but the code can also be implemented in 3D with spherical cells. For cell-cell interactions, we choose a Hertzian contact mechanism in which a short-range attraction force simulates cell adhesion, and an elastic repulsive force mimics the cell-cell repulsion (see appendix for details). We performed control simulations in which we confirmed that the results did not depend on the strength of the attraction or repulsion or the functional form of the repulsion: The cells need merely be sufficiently repulsive that the system remains approximately incompressible and sufficiently adhesive that the tissue does not tear. We implement our simulations in an over-damped setting with a cell-substrate friction that is proportional to the cell area: larger cells experience higher friction.

Our simulations are mainly performed in channels resembling a strip of cells with periodic boundaries in the y-direction, expanding in the +/- x-direction. Periodic boundaries strictly equate to a cylindrical shell of cells but likely provide an excellent approximation to a planar channel, provided that the walls of the channel themselves do not provide significant friction. The symmetry direction, along $x$, has an open boundary, while the $y$-direction is given a periodic boundary condition in lieu of explicit walls to the channel. Simulations begin with one cell with an unstressed 2D volume (area) $v_1=5\Vs$, consistent with the nominal average volume at a corresponding growth rate of $c=(v_1-1)\tau_v+\log2$ (see section \ref{continuumtheory} for derivation. This cell has a random initial phase, sampled uniformly from $\theta\in[0,2\pi]$\footnote{To avoid artificial synchronization in our particle based simulations we introduce some noise by sampling the division period for each cell, assigned at birth, from a normal distribution with mean $\taudiv$ and standard deviation $\taudiv/10$.} and is introduced into the centre of the channel.  It then divides after time $t\sim\taudiv$. As the cells repeatedly divide the colony expands roughly as a circle until it spans across the channel width, see Fig.~\ref{fig:method} (b)-(d); (e) shows how the colony then develops two fronts, moving with speed $s$ in the $\pm x$-directions. Unless specified otherwise simulations are performed at $\tau_v=\tau_r=\taudiv$, and with a dimensionless surface friction, relating force per cell area to sliding velocity, $\zeta=0.01$ solved using a timestep $\Delta t=10^{-6}$.

The pressure measurement of each cell is calculated by the virial stress tensor method \cite{zausch2009build}, equivalent to summing all the inward forces and dividing by the cell area (perimeter). For more detail on methods see appendix \ref{app:simu}.

\subsection{Continuum theory}\label{continuumtheory}
To gain analytic insight we develop a continuum theory of this cellular material, noting that continuum models underlie many of the recent physics-based models of cell cultures and tissues reviewed in the introduction. 

Cells grow in volume during the interdivision process, as given by \cref{Vdot}, while the cell division process always acts to reduce the volume of each cell: at a division event the cell volume halves. We seek to approximate this by a process that is continuous in time. The effect of division on the mean cell volume can be incorporated into a revised version of \cref{Vdot}, where we now write the division-adjusted mean volume as $v$, to distinguish it from the inter-divisional volume $V$,
\begin{equation}\frac{\dot v}v= r(c-\log 2) -\frac1{\tau_v}(v-1)\label{Vdotwithdivision}\end{equation}
The new term $-r\log 2$ provides a mean field approximation for the reduction in mean volume due to cell division. To see this note that $\frac{\dot v}v= -r\log 2$ has solution $v\propto \left(1/2\right)^{rt}$ with $1/r$ the time between volume-halving division events in units of $\taudiv$.

The mean cell volume at constant $r$ is then given by the solution $\dot v=0$, hence $v=1+\tau_v(c-\log 2)r$. For an unstressed cell cycle $r=1$ this defines a relationship between the unstressed division-adjusted mean volume $v_1$ and $c$
\begin{equation}v_1=1+\tau_v(c-\log 2) \label{V1def}\end{equation}
so that \cref{Vdotwithdivision} becomes
\begin{equation}\tau_v\frac{\dot v}v= r(v_1-1)-(v-1)\label{vdot}\end{equation}
Eq.~\ref{V1def} is also used  in the simulations to relate a nominal unstressed division-adjusted mean volume $v_1$ to growth rate $c$.

\subsubsection{Continuity equation}

We seek a continuity equation to establish the velocity $\u$ of the growing cell culture, treated as a continuum. To achieve this we exploit Gauss' theorem
$\int \u\cdot{\bf dS}=\int \nabla\cdot\u\>dV$
evaluated around an infinitesimal volume $dV$ containing $dn=dV/v$ cells of volume $v$. The total rate of change of volume is the integrated outward flux of cells $\int {\bf u}\cdot{\bf dS}=\frac{d}{dt}(v\>dn)$, hence $\nabla\cdot\u\>v\>dn = \frac{d}{dt}(v\>dn)=dn \dot v+v\dot dn$.
The change in the number of cells follows exponential growth according to $\dot dn= \log 2r\>dn$ in the mean field, while the average cell volume follows \cref{vdot}. Combining these results and dividing by $v\>dn$ we obtain the approximate relationship
\begin{equation}\nabla\cdot\u= \frac{\dot v}v +\log 2r=cr-(v-1)/\tau_v\label{divu}\end{equation}
with $c=\log 2+(v_1-1)/\tau_v$ from \cref{V1def}.

\subsubsection{Stokes-Darcy stress balance}
Cells are assumed to experience friction with a substrate with a friction coefficient\footnote{$\tilde\zeta=\zeta/p_r$, dropping the tilde; similarly for $\eta$ and $\xi$}  $\zeta$. In general there are also internal shear stresses, controlled by viscosities $\eta$ and $\xi$, leading to the stress balance equation
\begin{equation}\eta \laplacian \u +\xi \nabla\> \nabla\cdot\u-\zeta \u=\nabla p\label{stokes}
\end{equation}
\cref{divu,stokes} are analogous to Stokes equations for a low Reynolds number fluid but here incorporate the role of cell division and a Darcy-like frictional force, characterised by $\zeta$. The term in \cref{stokes} involving \hbox{$\nabla\cdot\u$}, with prefactor $\xi$ depending on the bulk viscosity, usually vanishes for incompressible fluids \cite{Landau1987Fluid}. In what follows, we set $\eta=\xi=0$, assuming that the growth is substrate-friction dominated.
\subsubsection{Advected description}

In the presence of the flow field $\u$ the lab-frame cell activity $r$ and volume $v$ are given by the advected forms of \cref{rdot} and \cref{vdot} 
\begin{equation}\frac{Dr}{Dt} r = \frac{\partial r}{\partial t} +\u\cdot \nabla r =  \frac{r}{\tau_r}\left(1-r - p\right)\label{rdotadvected}
\end{equation}
\begin{equation}\frac{D}{Dt} v = \frac{\partial v}{\partial t} + \u\cdot \nabla v  =  \frac{v}{\tau_v}\left(r(v_1-1)-(v-1)\right)\label{vdotadvected}
\end{equation}

In order to establish the dynamics we simultaneously solve for $\u$, $r$, $v$ and $p$ using \cref{divu,stokes,rdotadvected,vdotadvected}

\subsubsection{Channel flow (1D)}
For  consistency with the particle based simulations we assume $\eta=\xi=0$, noting that symmetry dictates that average motion is along the $x$-direction, i.e. $\u=u\hat{\bf x}$ and so shear stresses are  absent in a coarse-grained theory. We seek solutions in which there is a steady-state front of dividing cells moving with constant speed $s$. It is convenient to transform to the co-moving frame using $z=st-x$, focussing on the right-moving front where (the arbitrary zero of time is chosen so that) the front is at $z=0$ and the colony populates the space $z>0$. 
The partial derivatives transform as $\nabla=\frac{\partial}{\partial x}\to-\frac{\partial}{\partial z}$ and $\frac{\partial}{\partial t}\to s\frac{\partial}{\partial z}$, where we use a prime ($'$) to denote $\frac{\partial}{\partial z}$ in what follows. Hence \cref{divu,stokes,rdotadvected,vdotadvected} become
\begin{equation}-u'=c r-(v-1)/\tau_v\label{divu1d}\end{equation}
\begin{equation}p'=\zeta u\label{stokes1d}\end{equation}
\begin{equation}(s-u)r'=  \frac{r}{\tau_r}\left(1-r - p\right)\label{rdotadvected1d}\end{equation}
\begin{equation}(s-u)v'= \frac{v}{\tau_v}\left(r(v_1-1)-(v-1)\right)\label{vdot1d}\end{equation}

With boundary conditions
\begin{equation}r(0)=1\quad v(0)=v_1\quad p(0)=0\quad u(\infty)=0\label{BCs}\end{equation}
with $s=u(0)$ shorthand for the front speed, to be determined. We solve \cref{divu,stokes,rdotadvected,vdotadvected,BCs} numerically, as outlined in \cref{app:shooting}.
We can compute the corresponding boundary condition on the derivative of $r$ from \cref{rdotadvected1d} according to
\begin{equation}r'(0)=\lim_{z\to0}\frac{r(1-r - p)}{\tau_r(s-u)}\label{BCrprimea}\end{equation}
By L'Hopital's rule and using the boundary conditions \cref{BCs} this is
\begin{equation}r'(0)=\lim_{z\to0}\frac{\frac{\partial}{\partial z}(r(1-r - p))}{\frac{\partial}{\partial z}(\tau_r(s-u))}=\frac{-r'(0)-\zeta s}{\tau_r\log 2}\label{BCrprimeb}\end{equation}
Hence the (redundant) boundary condition
\begin{equation}r'(0)=-\frac{\zeta s}{1+\tau_r\log 2}\label{BCrdot}\end{equation}
Similarly for the derivative of $v$ from \cref{vdot1d}
\begin{equation}v'(0)=\lim_{z\to0}\frac{v\left(r(v_1-1)-(v-1)\right)}{\tau_v(s-u)}\label{BCvprimea}\end{equation}
Again using L'Hopital's rule and the boundary conditions \cref{BCs} we find
\begin{equation}v'(0)=\frac{v_1\left(r'(0)(v_1-1)-v'(0)\right)}{\tau_v\log 2}\label{BCvprimeb}\end{equation}
leading to
\begin{equation}v'(0)=-\frac{v_1(v_1-1)\zeta s}{\left(1+\tau_r\log 2\right)\left(v_1+\tau_v\log 2\right)}\label{BCvdot}\end{equation}

\subsubsection{Scaling relations}
In order to better understand the structure of the growth front we can compute characteristic lengthscales, at least at the scaling level, by analysing the rate of change of the mechanochemical variables near the front.
\begin{enumerate}
\item The pressure at the front obeys $p'(0)=\zeta s$ from \cref{stokes1d,BCs}. A scaling estimate of the distance behind the front at which the pressure reaches the cell cycle reference (stalling) pressure is $R_p=1/p'(0)$, hence 
\begin{equation}R_p=1/(\zeta s)\label{Rp}\end{equation}
\item The cell cycle activity obeys $r'(0)$ from \cref{BCrdot}. A scaling estimate of the distance behind the front that the cell cycle will stall to $r=0$ is $R_r=1/|r'(0)|$, hence
\begin{equation}R_r=\frac{1+\tau_r\log 2}{\zeta s}\label{Rr}\end{equation}
\item The cell volume obeys $v'(0)$ from \cref{BCvdot}. A scaling estimate of the distance behind the front at which the volume reaches the quiescent volume is $R_v=(v_1-1)/|v'(0)|$ hence
\begin{equation}R_v=\frac{\left(1+\tau_r\log 2\right)\left(v_1+\tau_v\log 2\right)}{\zeta s}\label{Rv}\end{equation}

\item The local lab-frame cell velocity obeys $u'(0)=-\log 2$ from \cref{divu1d,BCs}. A scaling estimate of the distance behind the front at which the cells become immobile is $R_u=s/|u'(0)|$, hence 
\begin{equation}R_u=\frac{s}{\log 2}\label{Ru}\end{equation}
Where $s$ can be calculated self-consistently or estimated by setting $R_u\sim R_p$ (say), revealing the scaling relations (that then holds for all variables)
\begin{equation}R\sim1/\sqrt{\zeta}\label{Rs}\end{equation}
and
\begin{equation}s\sim1/\sqrt{\zeta}\label{scalings}\end{equation}
\end{enumerate}

\section{Results} \label{sec:result}
\subsection{Substrate friction}
The friction forces experienced by cell colonies expanding along a 2D channel can be traced primarily to transient adhesion with the substrate  \cite{volfson2008biomechanical,mather2010streaming,boyer2011buckling,marinkovic2019microfluidic}.
While it remains difficult to measure the absolute substrate friction it is more feasible to make controlled variations, e.g. by preparing surfaces with different area fractions of surface coatings. Such an approach may allow a test of \cref{scalings} and ultimately may give a way to measure the substrate friction indirectly, e.g. by observing the spreading speed of pre-calibrated cell lines. 

\begin{figure}
	\centering
	\includegraphics[width=86mm]{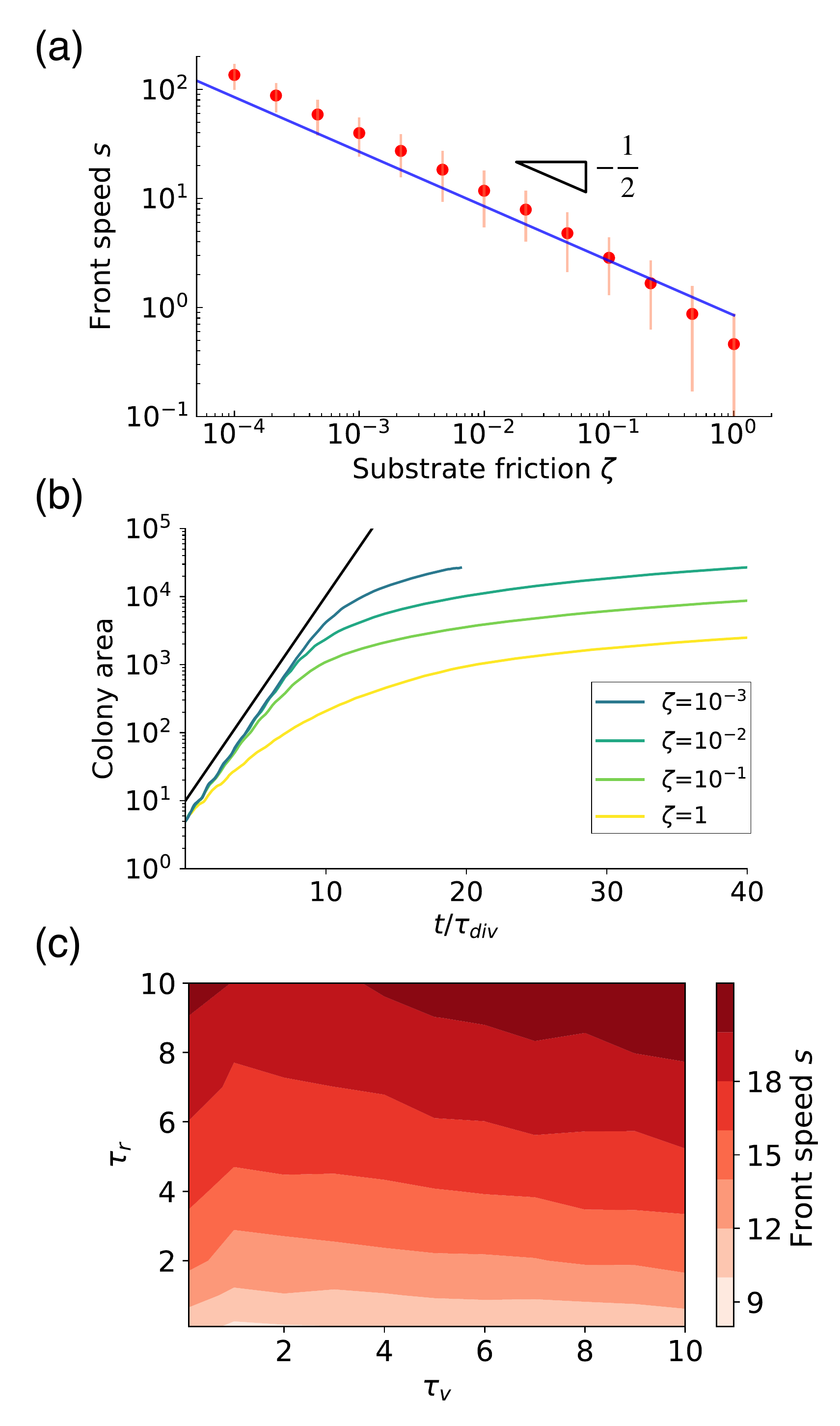}
	\caption{The front speed of an expanding colony scales like the inverse square-root of the substrate friction. (a) The dimensionless front speed $s$ is shown as a function of dimensionless substrate friction $\zeta$: Red points, and one standard-deviation error bars, are from particle based simulations, see appendix \ref{app:simu} for details, while the solid blue line is from the continuum theory via numerical solution of \cref{divu1d,stokes1d,rdotadvected1d,vdot1d} subject to boundary conditions (\ref{BCs}). (b) The dimensionless colony area first grows exponentially (solid black line shows exponential trend) becoming linear in time after steady growth fronts emerge. Each line represents the mean value of 5 independent simulations under each $\zeta$ condition. (c) At fixed surface friction $\zeta=0.01$ the front speed also depends on the dimensionless characteristic cell cycle times $\tau_r$ and, to a lesser extent, $\tau_v$.}
	\label{fig:substrate}
\end{figure}

Fig~\ref{fig:substrate}a) shows the relationship between front velocity and surface friction, revealing fully quantitative agreement between our particle-based simulations and continuum theory and confirming the scaling result \cref{scalings}. Fig~\ref{fig:substrate} also shows (b) the crossover from exponential growth to constant front speed and (c) an example of how the speed also depends on the cell cycle parameters.

\subsection{Colony structure}

\begin{figure*}
	\centering
	\includegraphics[width=2\columnwidth]{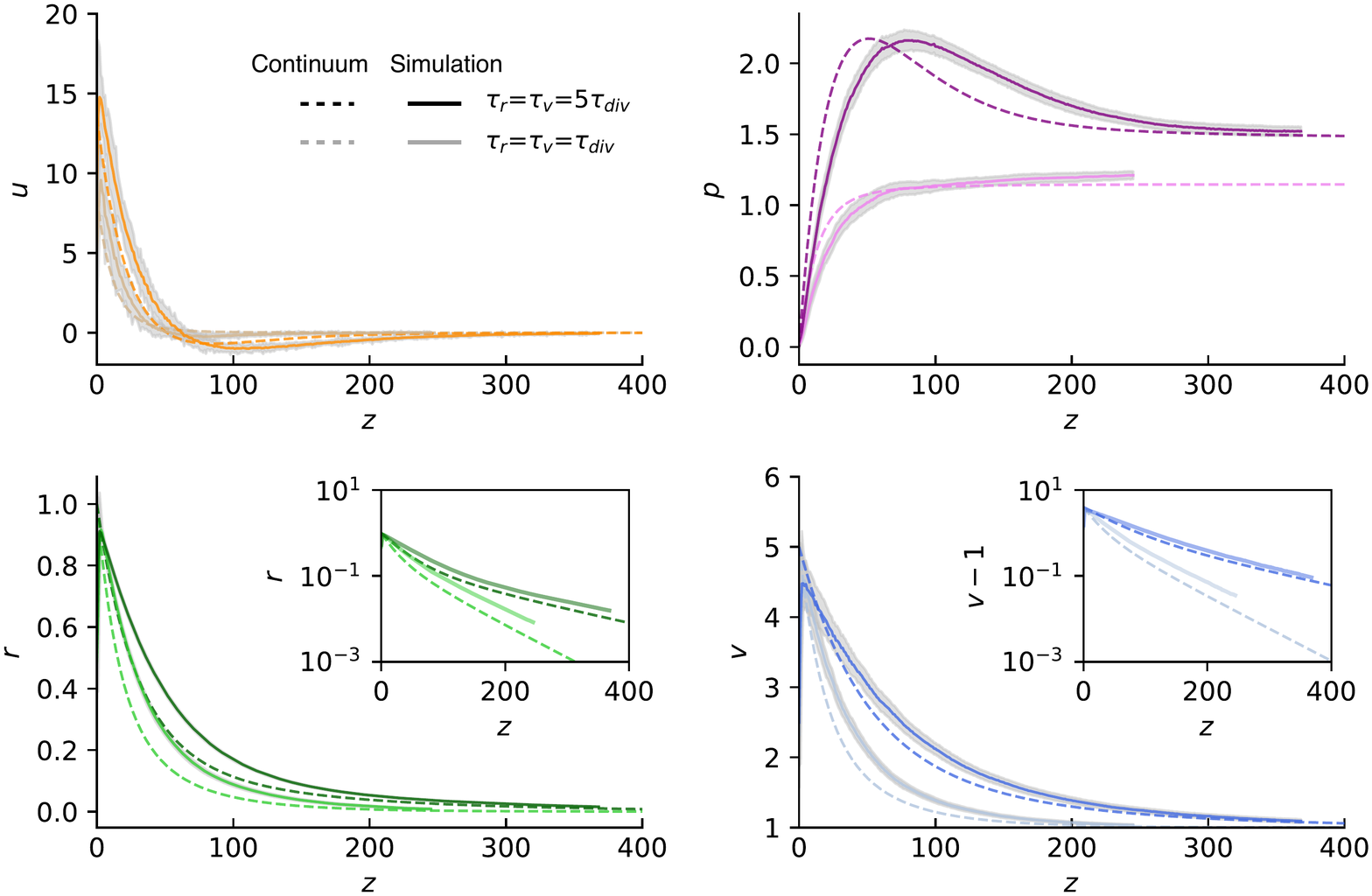}
	\caption{The mechanochemical variables exhibit different profiles behind the moving growing front with the simulations and continuum theory in semi-quantitative agreement. Shown are steady state average values of (a) lab-frame cell velocity $u$. (b) local cell pressure $p$, (c) cell cycle activity $r$, and (d) 2D cell volume (area) $v$, measured a distance $z$ behind the leading edge of the growing colony.  The solid lines are from particle based simulations with $\tau_r$=$\tau_v$=5$\taudiv$ (darker) and $\tau_r$=$\tau_v$=$\tau_{div}$ (lighter) (one standard deviation shown in grey). The dashed lines show the continuum solution. Insets in (c) and (d) reveal exponential decay deep in the bulk using a log-linear scale, see appendix \ref{app:expansion} for an analysis of these exponents.}
	 \label{fig:cmp}
\end{figure*}

During colony expansion, the organisational structure that emerges behind the growth front depends on the mechanochemical properties of individual cells, including the parameters controlling their cell cycle. We examine the steady state distributions of cell area, outward velocity, pressure, and cell cycle activity in a frame of reference co-moving with the front in which cells are active, growing and moving near the front and quiescent and stationary deep in the bulk, see Fig~\ref{fig:cmp} and Supplemental Material at [URL will be inserted by publisher] for movies of the growth process at various surface frictions and cell cycle times.

Fig~\ref{fig:cmp} (a) shows that the stochastic variation in the velocity is larger than in the other variables. This may be associated with the relatively large velocities that are realised, immediately post-division. Two daughter cells have large overlap immediately after division, and the repulsive force drives rapid separation. Although the post-division velocity is large this has almost no effect on the average velocity: the daughters separate in opposite directions.
We also predict the emergence of cells for which the sign of the velocity is reversed (located around $z\approx 100$ for these parameter values). This indicates that cells can sometimes move inward, away from the front, and is a result of slow volume loss of the cells approaching quiescence ``sucking'' cells in this region away from the front. This phenomenon can only be realised within relatively sophisticated models in which delayed volume loss is encoded into the cellular response. Similar negative velocities have been observed in experiments on multicellular spheres  \cite{delarue2013mechanical}. 

Fig~\ref{fig:cmp} (b) shows the pressure profile. Two features are worth highlighting. Firstly, the pressure in the bulk is systematically elevated slightly above $p_r$ ($p=1$ in dimensionless units). This is due to the controlled loss of division activity, which means some cells continue to divide even when the material is near the stalling pressure, resulting in an overshoot. {A consequence of the overshoot of the pressure is that the gradient of the pressure reverses its sign in the colony interior. This would provide a signature of contractility and is due to the shrinkage of the cells in this region.} Secondly, two types of pressure distribution can arise. When the cell cycle volume adaptation time $\tau_v$ is short enough the pressure monotonously increases from the edge toward the bulk, stabilising near $p_r$. Conversely, when $\tau_v$ is large the pressure is non-monotonic, exhibiting a pressure peak near the front, with the pressure decreasing further into the bulk. This is a consequence of relatively rapid volume loss of cells behind the growing front. Both of these features are quite generic. There are hints that a pressure peak might also be observed in experiments on multicellular spheroids~\cite{dolega2017cell}, although experimental measurements of pressure remain challenging. 

Fig~\ref{fig:cmp} (c) and (d) shows that both the cell cycle activity and cell volume decrease monotonically, approaching single exponential decay deep in the bulk (in the case of the volume decaying to the quiescent value $v=1$). Similar exponential decay emerges for $u$ and $p$ but this is not shown because of minor complications: the velocity can go negative and the bulk pressure is not fixed, see appendix \ref{app:expansion} for details. 

For all mechanochemical variables there is good semi-quantitative agreement between the continuum theory and the particle based simulations.

\begin{figure*}
	\centering
	\includegraphics[width=2\columnwidth]{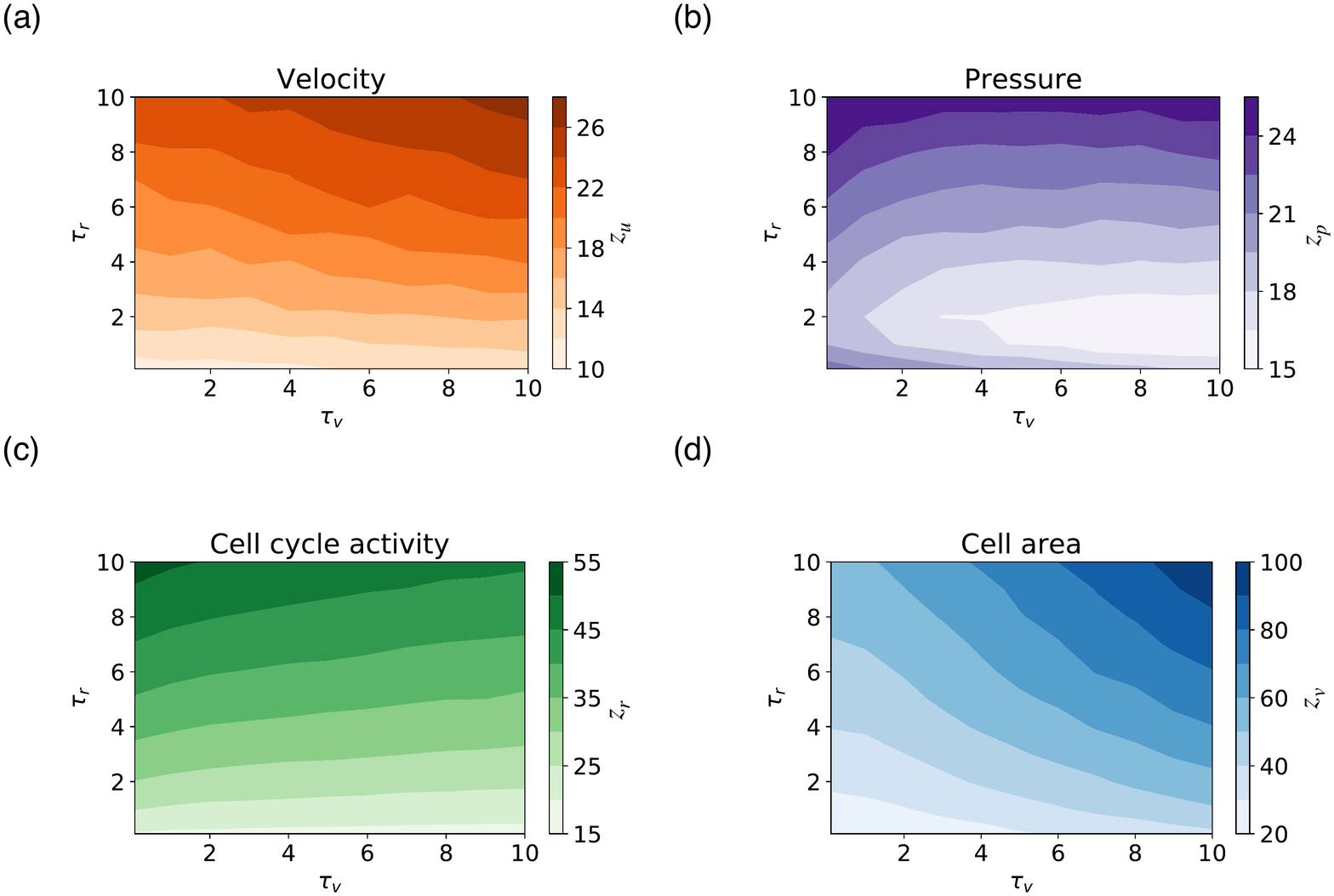}
	\caption{The dimensionless half-decay lengths for velocity $u$, pressure $p$, cell cycle activity $r$ and cell area (2D volume) $v$, near the growth front depend differently on the two dimensionless cell cycle times $\tau_r$ and $\tau_v$.}
	\label{fig:tau2decay}
\end{figure*}

\subsection{Cell cycle control of the growth front}
\begin{figure}
	\centering
	\includegraphics[width=86mm]{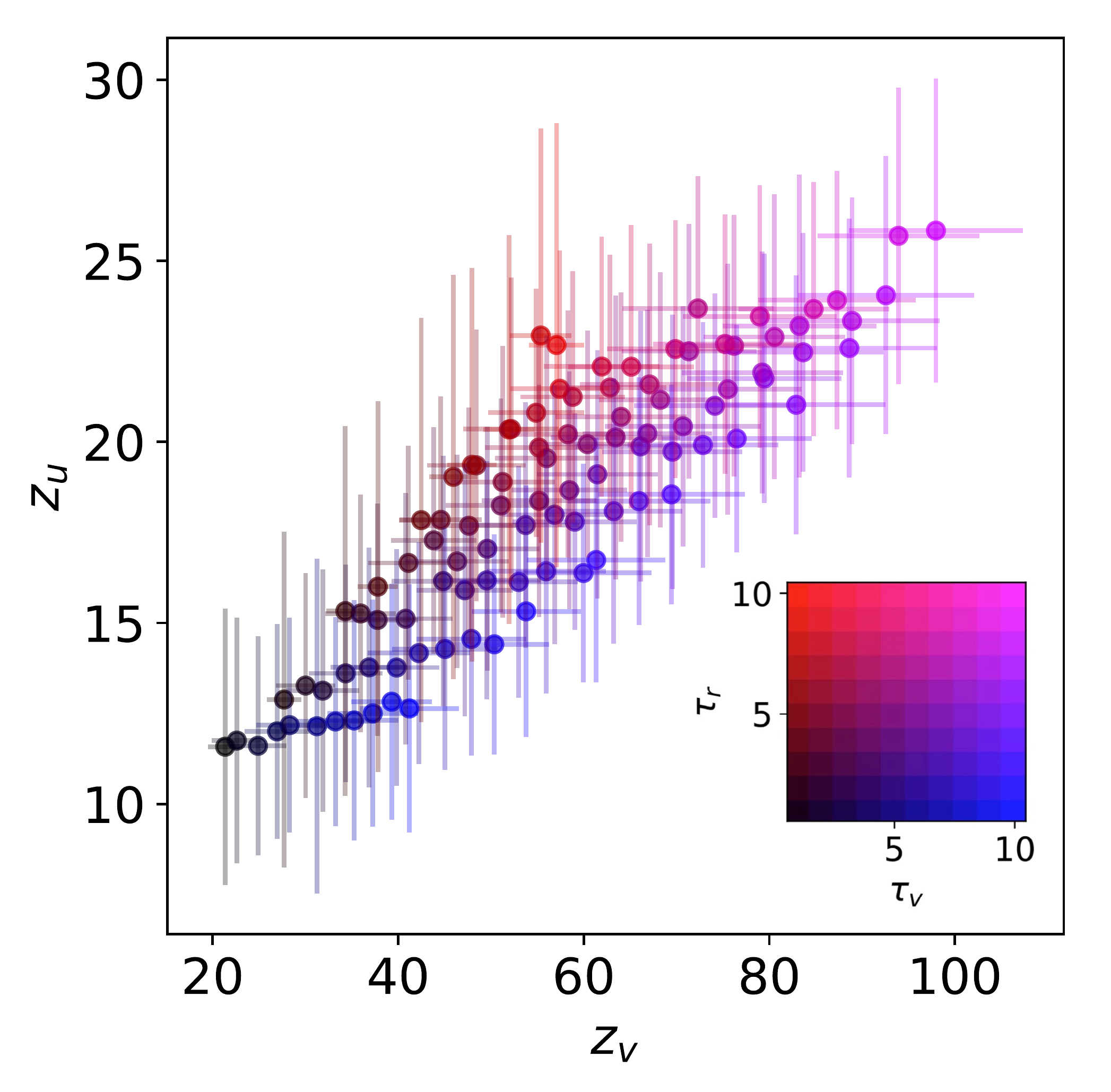}
	\caption{The cell cycle times $\tau_r$ and $\tau_v$ can be inferred from the half-decay lengths $z_v$ and $z_u$, respectively describing the distance over which volume and velocity drop to half their values at the front. The color of each point shows the value of $\tau_r$ and $\tau_v$ encoded into the simulations, see inset color map, while the position of the data point gives the different half-decay length combinations realised.}
	\label{fig:decay2tau}
\end{figure}

The cell cycle parameters $\tau_v$ and $\tau_r$ affect the physical distribution of the mechanochemical variables across the growth front. In order to analyze this we define a half-decay length for each variable written as $z_u,\>z_p,\>z_r,\>z_v$ respectively. These are defined as the smallest root of the following relations: For $r$ and $v$ we use $r(z_r)=0.5$ and $v(z_v)=(v_1-1)/2$; for $p$ and $u$ we use $p(z_p)=(p_{\rm max}-p_{\rm min})/2$, involving the empirically determined maximum and minimum values of pressure, similarly for $u$. See appendix~\ref{app:simu} for details. Fig 5 shows the relationship between half-decay lengths with the two cell cycle times $\tau_r$ and $\tau_v$ of velocity $u$, pressure $p$, cell cycle activity $r$ and cell area $v$ respectively. This relationship can be inverted to relate the experimentally observable half-decay lengths to the underlying cell cycle parameters, e.g. $\tau_r$ and $\tau_v$, see Fig~\ref{fig:decay2tau}. We choose to focus on the volume and velocity decay lengths here because they can be more easily visualised. 

\subsection{The cell cycle controls features inaccessible to purely ``physics based'' models}
In this section we review how the incorporation of  a stylised cell cycle introduces descriptive power not readily accessible to a phenomenological ``physics only'' model. 
\\

While modelling mechanical-feedback in cell dynamics is not new it is known that cellular behavior, very generally, is controlled by the cell cycle, a profoundly out-of-equilibrium chemical oscillator/sensor {\cite{alberts2015essential,matson2017cell}, see also the discussion in the introduction.} 
This regulates the cell proliferation state and {is involved in} other decisions the cell makes about its development~{\cite{benham2015mechanical}} except possibly under the most extreme stress when there is catastrophic (genuine physical) failure. 
 {A common understanding} in cell biology is that the cell cycle is the master controller with physical forces providing input to this controller {\cite{golias2004cell,petridou2017,Uroz2018,Radmaneshfar2013,Nam2019}}. 
This controller can also be manipulated in numerous other ways, e.g. temperature, chemical composition and genetic manipulation. It becomes increasingly contrived and clumsy to incorporate such factors into models that start from a purely physical description but relatively straightforward with (refinements to) models that incorporate a cell cycle explicitly. For these reasons we believe that it is important that models are able to reflect the underlying regulatory mechanisms of cell biology. However, the way in which a phenomenological ``physics only'' model differs from one with an explicit cell cycle regulation deserves to be addressed directly.  The most obvious signature of this difference here is the fact our results depend on the timescales $\tau_r$ and $\tau_v$ that enter through the cell-cycle \cref{rdot,Vdot}. These affect the emergent physical behaviour, including the front speed, velocity distribution, volume distribution, activity distribution and pressure distribution, as shown in Figs~\ref{fig:substrate}, \ref{fig:cmp}, \ref{fig:tau2decay} and \ref{fig:decay2tau}. Some of the qualitative features that are controlled by these parameters include the pressure overshoot and velocity sign change that are shown in Fig~\ref{fig:cmp}.
\\ 

In order to seek to underline this we identify a limiting case of our model, corresponding  to a ``physics only'' limit. We take a fast response limit of our cell cycle oscillator in which $\tau_r=\tau_v=0$; neither parameter then appears explicitly. \Cref{rdot,vdot} then reduce to $r=1-p$ and $v=1+r(v_1-1)$, with $r$ and $v$ slave to the instantaneous $p$ and $r$, in the spirit of direct physical control. \cref{divu1d,stokes1d} then reduce to $u'=-c r$ and $p'=\zeta u$. Substituting the former into a derivative of the later we obtain $p''=c\zeta(p-1)$. Hence \hbox{$p=1-e^{-z/l}$} with $l=1/\sqrt{c\zeta}$. All other variables ($r,u,v$) are similarly mono-exponential, with the same lengthscale. This universal mono-exponential behaviour is clearly quite distinct from that arising under more general values of $\tau_r$ and $\tau_v$, as can be seen from Fig~\ref{fig:cmp}, e.g. the decay lengths vary and the pressure and velocity can be non-monotonic.

Finally, it is also well known that driven oscillators have distinct properties from non-oscillatory systems \cite{synchronisation}. Given that a genuine oscillator is present in our model one might also anticipate improved descriptive power, e.g. under conditions where the system is driven in a time-varying fashion with a refined $f(r,p,\theta)$, introduced to capture the phase (checkpoint) sensitivity of physical stimulus.

\begin{figure*}
	\centering
	\includegraphics[width=2\columnwidth]{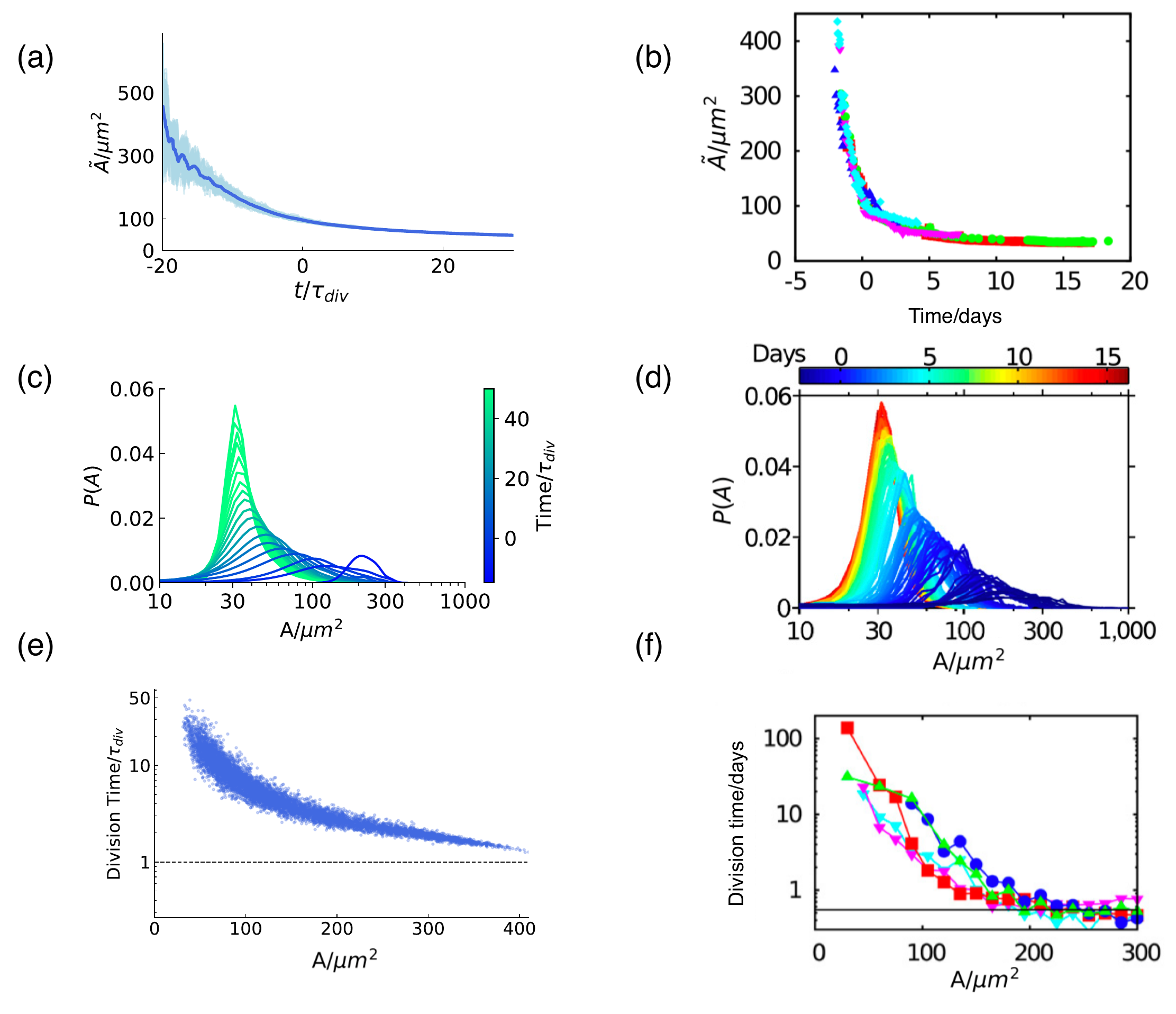}
	\caption{Cell area distributions in MDCK cells are reproduced in our simulations. The three panels on the left (a,c,e) are simulation results; the panels on the right (b,d,f) are experimental results on spreading MDCK cells, reproduced with permission from  
	\cite{puliafito2012collective}. Panels (a,b) show the evolution of colony-average median cell area  ${\tilde{A}}$ as a function of time. In panel (a) we set $t=0$ as the crossover from exponential to subexponential growth of the colony area, roughly 20$\taudiv$ after initialisation. In panel (b) time $t=0$ is measured from a ``morphological transition'' that is close to the crossover to subexponential growth, noting that substantial changes in cell thickness were reported before this time, see \cite{puliafito2012collective} for details. These thickness changes mean that comparison with quasi 2D simulations (a) only becomes appropriate for $t\gtrsim 0$. Panels (c,d) show the colony-wide probability distribution of individual cell area at different times (colors). Panels (e,f) show the distribution of the time between cell division events as a function of the premitotic area. In (e) every cell division event contributes a data point while in (f) data points with division times greater than 3 days are inferred indirectly. In (f) the five different datasets (colours) are from different experiments. See \cite{puliafito2012collective} for details. We choose simulation parameters to roughly match those of the MDCK cells: $\Vs=30\mu{\rm m}^2$, $v_1=15\Vs$ and $\taudiv=\tau_r=\tau_v=0.5$ day, see appendix \ref{parameters} for details.
	}
	\label{fig:areas}
\end{figure*}

\subsection{Dynamics of area distribution}\label{sec:areas}

We compare our model with experimental data for the growth of 2D sheets of MDCK cells \cite{puliafito2012collective}, see Fig~\ref{fig:areas}. 
Our simulations show broadly similar trends in cell area distributions and division rates to those observed in MDCK cells, noting that these cells are known to be motile and so this level of agreement is unexpected. Rather surprisingly, we were unable to locate suitable experimental data on non-motile cells in 2D that are in the non-exponential growth phase and not limited by other factors, e.g. nutrients. That we provide motivation for such experiments in the future should be considered an objective of this work.

Fig.~\ref{fig:areas} (b) shows the fastest division rate for the largest cells to be approximately $\taudiv=0.5$ day. Hence we take $\tau_r = \tau_v =\taudiv$ as nominal values for the cell cycle times and fix $\Vs=30\mu {\rm m}^2$ and $v_1=15\Vs$. 

In our simulations the cells adjust their size naturally according to internal cell cycle activity $r$ and hence local pressure $p$. The simulations are initialised with a single cell in the centre of 2D substrate, generating a roughly circular growth front for all times, see Supplemental Material at [URL will be inserted by publisher] for a movie of colony growth in this geometry. This is to align with the experiments and is different to the 2D channel considered in the rest of this study. As a steadily expanding colony emerges the cells in the bulk (interior) start switching to the quiescent state, with an area that approaches $\Vs$. As a result the median area of the colony decreases towards $\Vs$ in the late stage of simulations, as Fig.~\ref{fig:areas} (a) and (b) both show.  Time $t=0$ in panel (b) is measured from a ``morphological transition'', close to the crossover to subexponential growth. The experiments reported cells that that were thinner in the early stages of growth, with the cell thickness roughly constant once the median area fell below $100-150\mu{\rm m}^2$. This means the system only has a quasi 2D nature, with area a proxy for volume, after this point. Hence our theory can only reasonably be expected to apply in this regime.

We also record the distribution of cell areas at different times. Fig.~\ref{fig:areas} (c) and (d) both show the cell areas converging to a narrow, stationary size distribution centered around $30-35\mu {\rm m}^2$. Because of the way we have incorporated cell cycle activity into our model the division time for individual cells can vary enormously, depending on local pressure, as both Fig.~\ref{fig:areas} (e) and (f) show. This leads to a range of division times from $\taudiv$ up to $\sim 50\taudiv$ with large (unstressed) cells dividing faster than smaller cells. This degree of  agreement with the experimental data would not be produced by models that do not incorporate a similar mechanism to control division rate and size. 

\section{Discussion} \label{sec:discussion}
 We have developed a model for confluent cells that incorporates a stylised cell cycle, regulating division and cell size. This both incorporates and, indirectly generates, mechanical feedback: the former via pressure sensitivity of the cell cycle activity and the latter via the growth and division of cells, driving flows and generating dynamical stresses. Our motivation is to develop a minimal model of this feedback, noting that the cell cycle model can be made more sophisticated, e.g. by alternative choices of $f$ or $g$ in \cref{thetadotdimensional,rdotdimensional}; the incorporation of apoptosis, cell motility or the supply and removal of nutrients, oxygen and waste byproducts; extensions to 3D or heterogeneity, e.g. mimicking different tissue or cell types. We developed a particle-based simulation to study this model, complemented with a simple continuum theory that is found to be in broad agreement with the simulations. In this work we have focussed mainly on colony development in a quasi 2D channel with adjustable substrate friction, neglecting the role of active motility.  

This simple model may be useful as a reference tool to characterise substrate friction using pre-calibrated cell lines. We also studied the spatial structure of the growing front, where cells are proliferating and moving outwards. We focus on four fundamental variables: outward velocity, local pressure, cell cycle activity and cell area. Cells switch from proliferation to a quiescent state as they transition from the front into the interior (bulk). We propose a method to relate the parameters underlying the cell cycle model to experimental observables. We also compare our simulations with experiments on expanding colonies of MDCK cells, noting that these cells are motile, a feature absent in this version of our model. We find broad agreement for the area distribution and division rates, supporting the use of sizing and division mechanisms under the control of the cell-cycle. 

In summary, we hope this work has helped to establish that the cell cycle can play a non-trivial role in the physics of dividing cells and that this might inform future model development. {We hope the current paper motivates experimental work on growing cell colonies in 2D to isolate and further evaluate the role of the cell cycle in collective cell dynamics.} \\

\section*{Acknowledgements} \label{sec:acknowledgements}
We thank Carles Blanch-Mercader, Estelle Gauquelin, and John J. Molina for insightful discussions. J.L. thanks the Uehara Memorial Foundation for a fellowship. M.S.T. acknowledges the generous support of visiting fellowships from JSPS and the Leverhulme Trust and the kind hospitality of the Yamamoto group. This work is supported by the Grants-in-Aid for Scientific Research (JSPS KAKENHI) under Grants No. 20H00129 and 20H05619. The simulation videos were rendered with Ovito\cite{stukowski2009visualization}.

\appendix\label{sec:appendix}

\section{Simulation Details}\label{app:simu}
\subsection{Intercellular interaction}
Cells are modelled as soft spheres (disks in 2D) with the interaction between cells described by an elastic repulsive force and constant adhesive force per unit contact area. 
The repulsive force experienced by cell $i$ due to interactions with cell $j$ is assumed to follow Hertzian contact mechanics  \cite{schaller2005multicellular} according to
\begin{equation}
	{\vec F_{ij}^{\el}}(t)=\cfrac{ f^{\el}\>h_{ij}^{3/2}(t)\>{\vec n_{ij}} }{\sqrt { {1}/{R_{i}(t)}+ {1}/{R_{j}(t)}}}
\end{equation}
This force act along the unit vector ${\vec n_{ij}}$ pointing from the center of cell \textit{i} to the center of cell \textit{j}. All variables are dimensionless (see Table~\uppercase\expandafter{\romannumeral1}) and all lengths are measured in the quiescent cell size, $\sqrt\Vs$ in 2D. 

\begin{figure}
\begin{center}
	\includegraphics[width=86mm]{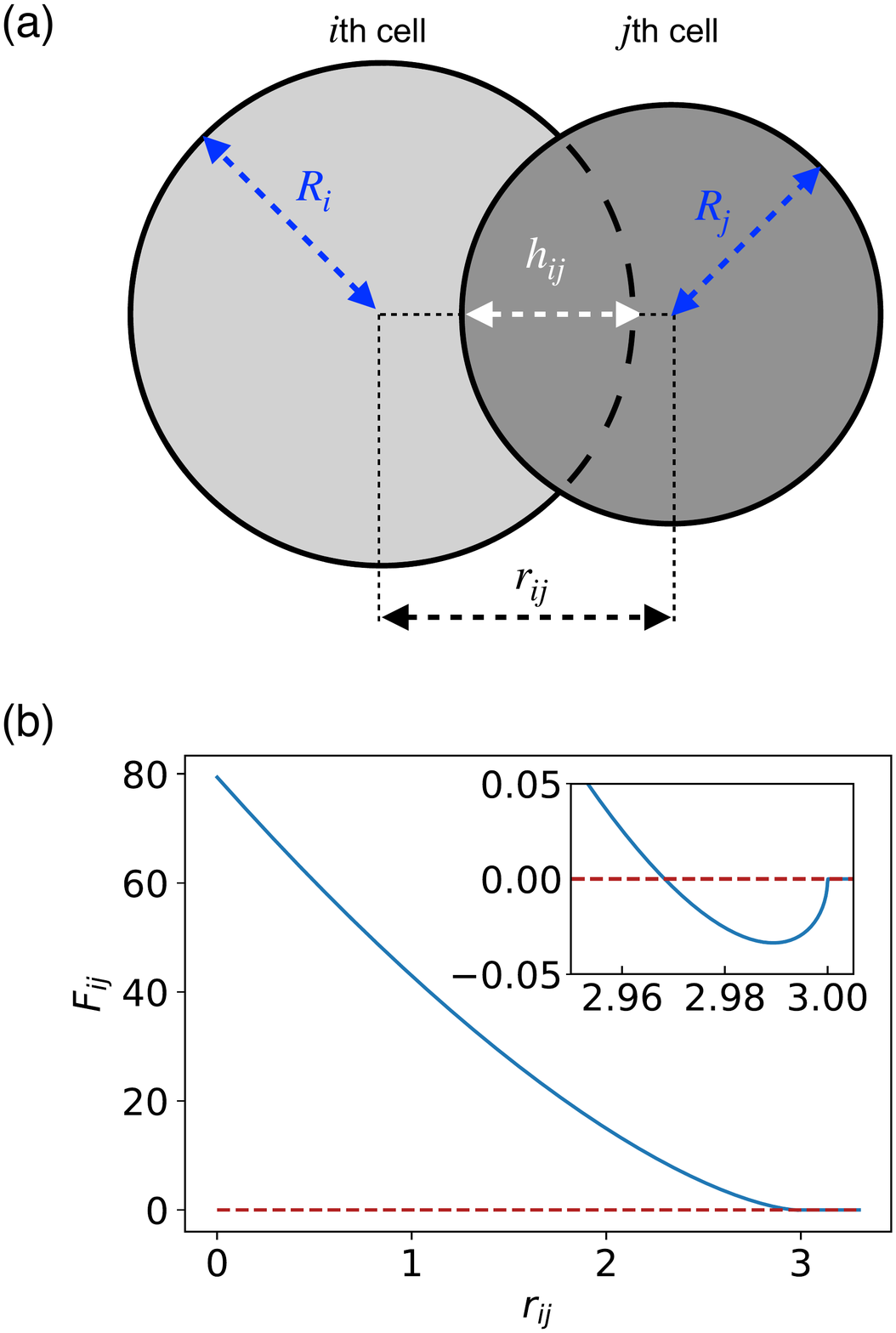}
	\caption{(a) Two contacting cells $i$ and $j$ with radii $R_i$ and $R_j$. The center-to-center distance between two cells is $r_{ij}=|R_i-R_j|$, and the overlap is $h_{ij}$, given in \cref{hij}. (b) The force on cell $i$ due to cell $j$, $F_{ij}$, is plotted as a function of the separation $r_{ij}$ with $R_i=R_j=1.5$. The inset reveals the short-range attractive force. $F_{ij}=0$ if there is no contact, i.e. $r_{ij}>R_i+R_j=3$ here.} 
	\label{fig:Force}
\end{center}
\end{figure}

The overlap of two cells $h_{ij}$ is defined in terms of their center-to-center distance $r_{ij}$, see Fig~\ref{fig:Force}. The force per area $f^{\el}$ for solid elastic Hertzian spheres can be identified with 
$f^{\el}=({4}/{3})/\left(({1-\nu_{i}^2})/{E_{i}}+({1-\nu_{j}^2})/{E_{j}}\right)$ 
with $R_i$, $E_i$ and $\nu_i$ the radius, elastic modulus and Poisson ratio of the $i^{\rm th}$ cell, respectively. Alternatively $f^{\el}$ can be treated merely as a single (here, constant for all $i,j$) adjustable prefactor, controlling repulsion. 
The dimensionless overlap between cells $i$ and $j$ is 
\begin{equation}
h_{ij}=\begin{cases}
R_i+R_j-r_{ij}& {\rm for}\quad r_{ij}<R_i+R_j\\
0 & {\rm for}\quad r_{ij}\ge R_i+R_j
\label{hij}
\end{cases}
\end{equation}
This vanishes for cells that don't overlap: such cells naturally have no interactions. The intercellular adhesion force between cells $i$ and $j$ follows from the approximation that receptor-ligand interactions scale with the dimensionless contact area $A_{ij}=A_{ji}$ 
\begin{equation}
		{\vec F_{ij}^{ad}}=-f^{ad} A_{ij}{\vec n_{ij}}
\end{equation}
with $f^{ad}$ a constant that can be related to a more microscopic model for receptor-ligand binding, if desired. In our 2D simulations this contact becomes a dimensionless length, defined as 
\begin{equation}
A_{ij}={\sqrt{(r_{ij}^2-(R_i-R_j)^2)((R_i+R_j)^2-r_{ij}^2)}}\Big/{r_{ij}}
\end{equation}

This gives a total force between cells $i$ and $j$ of $\vec F_{ij}=\vec F_{ij}^{\el} +\vec F_{ij}^{ad}$. Summing all interactions over $j$ gives the force on the $i^{\rm th}$ cell
\begin{equation}
	\vec F_i=\sum_{j} \vec F_{ij}
	\label{sumforces}
\end{equation}
Smoothly varying interactions of the kind chosen here are numerically convenient but the precise form of the interactions is unlikely to be important provided there is some weak attraction and strong enough repulsion to suppress excessively large cell indentations. 

\subsection{Pressure}
During the simulation the instantaneous dimensionless pressure on each cell, measured in units of $p_r$, is calculated by summing the scalar (inward-pointing) force and dividing by the 2D equivalent of the cell's surface area - its circumference, noting that the pressure can be negative if a cell is mainly experiencing attraction towards its neighbours. 
\begin{equation}
	p_i(t)=\sum_{j\neq i}\cfrac{-{\vec F_{ij}}\cdot{{\vec n_{ij}}}}{2\pi R_i(t)}
\end{equation}

\subsection{Half-decay lengths}
\begin{figure}
\begin{center}
	\includegraphics[width=86mm]{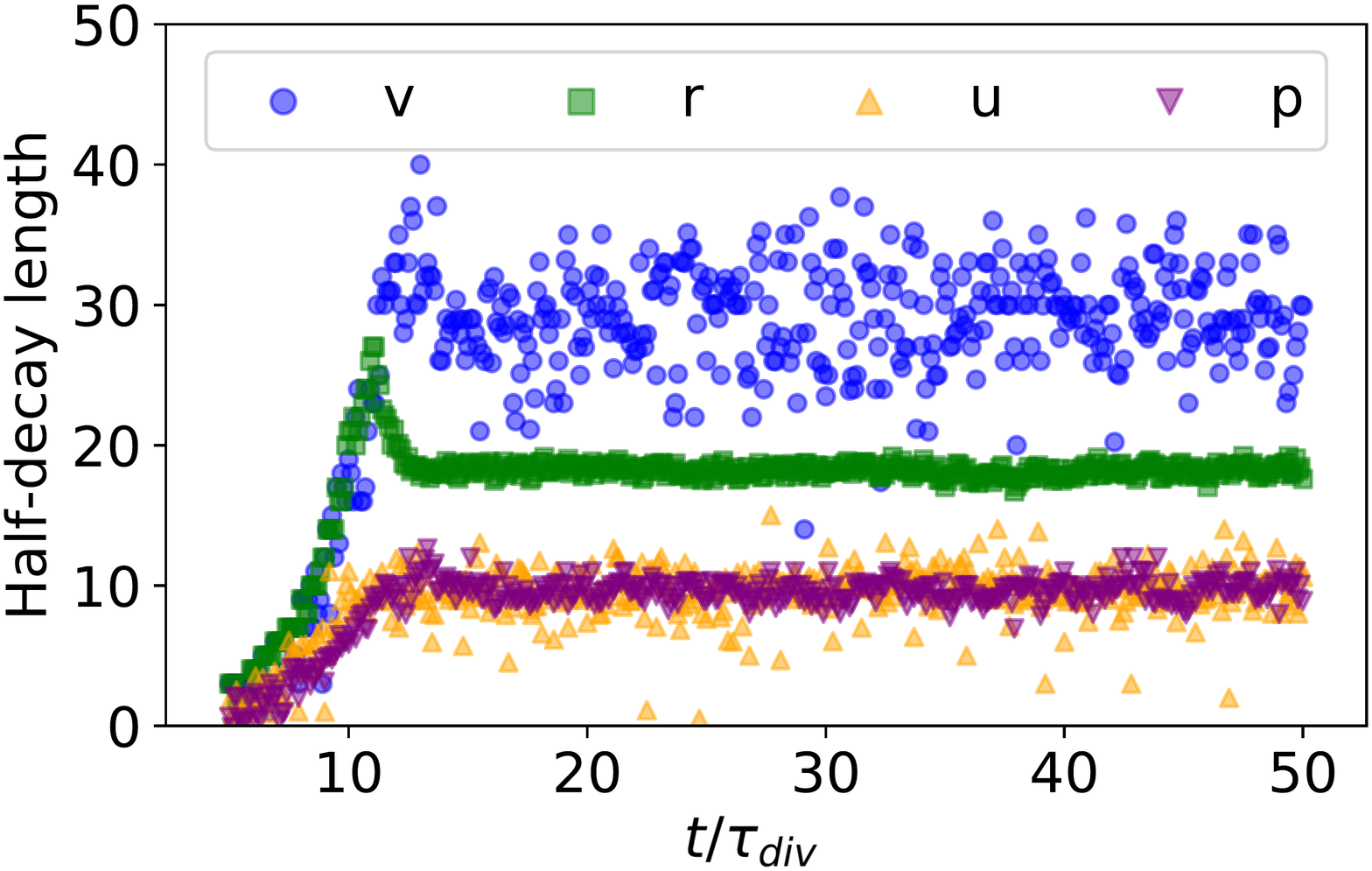}
	\caption{Time series of the instantaneous estimate of the dimensionless half-decay length for $\tau_r$=$\tau_v$=5$\tau_{div}$. The mean values are stable after $\sim20\tau_{div}$ and all results in the main text are reported in such a steady-state regime using improved averaging over all data in that regime.}
	\label{fig:DynamicHalfDecay}
\end{center}
\end{figure}

In order to estimate the half decay lengths described in the main text we process the simulation results using spatial binning along the channel direction and then use linear interpolation between the binned values. Before sampling, we ensure these have reached steady-state values, see Fig.~\ref{fig:DynamicHalfDecay}.

\subsection{Equations of motion}
The over-damped equation of motion for the position $\vec r_i$ of cell $i$ is
\begin{equation}
	\frac{d{\vec r_i}}{dt}=\frac{1}{\pi R_i^2\zeta}{\vec F}_i\label{eqofmotion}
\end{equation}
where the friction constant $\zeta$ is a friction per quiescent cell area, $\Vs$ in 2D, and so the factor $\cfrac{1}{\pi R_i^2}$ scales this to the current contact area of the $i^{\rm th}$ cell. 
Eq~\ref{eqofmotion} is integrated forward in time for each particle using a time step $\Delta t$ chosen to be small enough for good numerical stability and computational accuracy, see Table~\uppercase\expandafter{\romannumeral1}. 
\subsection{Model parameters and units}\label{parameters}
The standard values of all parameters, used in all simulations unless specified otherwise, are given in Table~\uppercase\expandafter{\romannumeral1}. 

\begin{table}[h!]\label{tableone}
\centering
\caption{Standard cell parameters}
\begin{tabular}{lcc}
\hline
\hline
Parameter                                             & Value                                                                                                          & Units \\ 
\hline
Channel width         & \begin{tabular}[c]{@{}l@{}}30\end{tabular} & $\sqrt{\Vs}$               \\
Substrate friction ($\zeta$)                  & 0.01                                                                                                                    & $\tau_{div}p_r/\Vs$                  \\
Time step ($\Delta t$)                        & 10$^{-6}$                                                                                                                   & $\tau_{div}$                  \\
Repulsive coefficient ($f^{rep}$)                         & 17.78                                                                                                                     & $p_r/\SqVs$                    \\
Adhesion coefficient ($f^{ad}$)                         & 0.2                                                                                                                     & $p_r$                    \\
Cell cycle activity recovery time ($\tau_r$) & 1                                                                                                                       & $\tau_{div}$                   \\
Volume recovery time ($\tau_v$)              & 1                                                                                                                       & $\tau_{div}$                   \\
Nominal unstressed cell area ($v_1$)                                & 5                                                                                                                       & $\Vs$              \\
\hline
\end{tabular}

\end{table}

\section{Continuum theory: numerics and asymptotics}

\subsection{Shooting method} \label{app:shooting}

The boundary value problem given by \cref{divu1d,stokes1d,rdotadvected1d,vdot1d} and boundary conditions
\cref{BCs} for $z>0$ are solved numerically as follows:  At the outset, the value of the front speed $s$ is unknown because it self-consistently depends on the full solution. However, we assert that the speed of the cells far away from the front must eventually decay to zero, $u(z)\to 0$ for $z$ large enough. The standard approach for such a system is the shooting method: By varying $s$ systematically, a front speed can be found for which the cell speed far away from the front correctly decays to 0. For any given $s$, the system becomes an initial value problem. This problem can then be solved by numerically integrating from the growing front, $z=0$, to a distance $z_{max}$ that is large enough for all quantities to have approximately decayed to their bulk values. This distance is typically of the order of many hundreds to thousands of cell diameters. We confirm afterwards that $z_{max}$ is much larger than all the decay lengths of the system. 
Since $s$ and $u(z_{max})$ are scalar, the front speed for which the bulk speed decays to zero can be found with a scalar root finder.

A slight obstacle is that the derivatives of $r$ and $v$ cannot be calculated numerically from \cref{rdotadvected1d,vdot1d} at $z=0$, since $(u(z=0)-s)=0$. We use \cref{BCrdot,BCvdot} instead.

\subsection{Asymptotic expansion around the bulk state}\label{app:expansion}

We study how the variables asymptotically approach the constant values characterising the tissue deep in the bulk, $u(z) \to 0$, $p(z) \to p_{bulk}$, $r(z) \to 0$, and $v(z) \to 1$, for $z\to \infty$, with $p_{bulk}$ some unknown pressure, see insets to Fig~\ref{fig:cmp}.
Expanding about the bulk state as $u(z) = \delta u(z)$, $p(z) = p_{bulk} + \delta p(z)$, $r(z) = \delta r(z)$, $v(z) = 1 + \delta v(z)$  the equations 
\cref{divu1d,stokes1d,rdotadvected1d,vdot1d} reduce to
\begin{align*}
  \delta u' &= -c\delta r + \frac{\delta v}{\tau_v}\\
  \delta p' &= \zeta \delta u\\
  \delta r' &= \frac{1}{s\tau_r}(1 - p_{bulk})\delta r\\
  \delta v' &= \frac{1}{s\tau_v} \left(\delta r (v_1 - 1) - \delta v\right).
\end{align*}
The system of equations for the perturbations can be written as a matrix equation
\begin{align*}
  \frac{\partial}{\partial z}
  \left (\begin{matrix}
    \delta u \\ \delta p \\ \delta r \\ \delta v
    \end{matrix}\right) = \left (\begin{matrix}
    0 & 0 & -c & 1/\tau_v \\ 
    \zeta   & 0 & 0 & 0         \\ 
    0 & 0 & (1 - p_{bulk})/s\tau_r & 0  \\ 
    0 & 0 & (v_1 - 1)/s\tau_v & -1/s\tau_v
    \end{matrix}\right)
    \left (\begin{matrix}
    \delta u \\ \delta p \\ \delta r \\ \delta v
    \end{matrix}\right)
\end{align*}
 The matrix has the repeated eigenvalue $e_0 = 0$ with eigenvector $\vec \psi_{0} = (0,1,0,0)$, the eigenvalue $e_1 = -1/(s\tau_v)$ with eigenvector $\vec \psi_1 = (-s, s^2 \zeta\tau_v, 0, 1)$, and the eigenvalue $e_2 = -(p_{bulk} - 1)/(s \tau_r)$ with eigenvector
\begin{equation*}
  \vec \psi_2 = \left( \begin{matrix}
    s\left[\frac{ \log (2) \kappa}{p_{bulk} -1} - 1\right], \\
    -\zeta \tau_r\frac{s^2}{p_{bulk} -1}\left[\frac{\log (2)\kappa}{p_{bulk} -1} - 1\right], \\
    \kappa/\tau_r, \\
    1
  \end{matrix}\right)
\end{equation*}
with $\kappa =  (\tau_r  - (p_{bulk} -1) \tau_v)/(v_1 -1)$. The solution can then be written as
\begin{align}
  \left (\begin{matrix}
    \delta u \\ \delta p \\ \delta r \\ \delta v
    \end{matrix}\right)
		= 
		A_1 \vec \psi_1 \exp\left[-\frac{1}{s\tau_v}z\right]+ A_2 \vec \psi_2 \exp\left[-\frac{p_{bulk} -1}{s\tau_r}z\right]
\end{align}
with two amplitudes $A_1$ and $A_2$ which have to be matched to the data.  Due to the repeated eigenvalue $0$, the solution would normally also contain a constant term and a term linear in $z$, but we can exclude that because of the condition that the solution must tend to zero in the bulk, $z \to \infty$. We observe that exponential decay with exponent $e_2 = -(p_{bulk} - 1)/(s \tau_r)$ tends to describe the data well, see insets of \cref{fig:cmp}. For all our data, we find that $e_1>e_2$, which implies that the asymptotic solution always becomes dominated by the latter exponent eventually. In practice, the bulk pressure has to be extracted from the numerical solution; we use the pressure for the largest simulated $z$ for each data set.  The fact that a single eigenvalue dominates deep in the bulk means that only a single parameter value, or combination thereof, could be inferred by fitting to corresponding data. This supports the focus on the neighbourhood of the growing front adopted in the main text: The front region provides for better model discrimination and parameter inference. 


\begin{thebibliography}{93}%
\makeatletter
\providecommand \@ifxundefined [1]{%
 \@ifx{#1\undefined}
}%
\providecommand \@ifnum [1]{%
 \ifnum #1\expandafter \@firstoftwo
 \else \expandafter \@secondoftwo
 \fi
}%
\providecommand \@ifx [1]{%
 \ifx #1\expandafter \@firstoftwo
 \else \expandafter \@secondoftwo
 \fi
}%
\providecommand \natexlab [1]{#1}%
\providecommand \enquote  [1]{``#1''}%
\providecommand \bibnamefont  [1]{#1}%
\providecommand \bibfnamefont [1]{#1}%
\providecommand \citenamefont [1]{#1}%
\providecommand \href@noop [0]{\@secondoftwo}%
\providecommand \href [0]{\begingroup \@sanitize@url \@href}%
\providecommand \@href[1]{\@@startlink{#1}\@@href}%
\providecommand \@@href[1]{\endgroup#1\@@endlink}%
\providecommand \@sanitize@url [0]{\catcode `\\12\catcode `\$12\catcode
  `\&12\catcode `\#12\catcode `\^12\catcode `\_12\catcode `\%12\relax}%
\providecommand \@@startlink[1]{}%
\providecommand \@@endlink[0]{}%
\providecommand \url  [0]{\begingroup\@sanitize@url \@url }%
\providecommand \@url [1]{\endgroup\@href {#1}{\urlprefix }}%
\providecommand \urlprefix  [0]{URL }%
\providecommand \Eprint [0]{\href }%
\providecommand \doibase [0]{https://doi.org/}%
\providecommand \selectlanguage [0]{\@gobble}%
\providecommand \bibinfo  [0]{\@secondoftwo}%
\providecommand \bibfield  [0]{\@secondoftwo}%
\providecommand \translation [1]{[#1]}%
\providecommand \BibitemOpen [0]{}%
\providecommand \bibitemStop [0]{}%
\providecommand \bibitemNoStop [0]{.\EOS\space}%
\providecommand \EOS [0]{\spacefactor3000\relax}%
\providecommand \BibitemShut  [1]{\csname bibitem#1\endcsname}%
\let\auto@bib@innerbib\@empty
\bibitem [{\citenamefont {Friedl}\ and\ \citenamefont
  {Gilmour}(2009)}]{friedl2009collective}%
  \BibitemOpen
  \bibfield  {author} {\bibinfo {author} {\bibfnamefont {P.}~\bibnamefont
  {Friedl}}\ and\ \bibinfo {author} {\bibfnamefont {D.}~\bibnamefont
  {Gilmour}},\ }\bibfield  {title} {\bibinfo {title} {Collective cell migration
  in morphogenesis, regeneration and cancer},\ }\href
  {https://doi.org/doi:10.1038/nrm2720} {\bibfield  {journal} {\bibinfo
  {journal} {Nat. Rev. Mol.}\ }\textbf {\bibinfo {volume} {10}},\ \bibinfo
  {pages} {445} (\bibinfo {year} {2009})}\BibitemShut {NoStop}%
\bibitem [{\citenamefont {Sadati}\ \emph {et~al.}(2013)\citenamefont {Sadati},
  \citenamefont {Qazvini}, \citenamefont {Krishnan}, \citenamefont {Park},\
  and\ \citenamefont {Fredberg}}]{sadati2013collective}%
  \BibitemOpen
  \bibfield  {author} {\bibinfo {author} {\bibfnamefont {M.}~\bibnamefont
  {Sadati}}, \bibinfo {author} {\bibfnamefont {N.~T.}\ \bibnamefont {Qazvini}},
  \bibinfo {author} {\bibfnamefont {R.}~\bibnamefont {Krishnan}}, \bibinfo
  {author} {\bibfnamefont {C.~Y.}\ \bibnamefont {Park}},\ and\ \bibinfo
  {author} {\bibfnamefont {J.~J.}\ \bibnamefont {Fredberg}},\ }\bibfield
  {title} {\bibinfo {title} {Collective migration and cell jamming},\ }\href
  {https://doi.org/10.1016/j.diff.2013.02.005} {\bibfield  {journal} {\bibinfo
  {journal} {Differentiation}\ }\textbf {\bibinfo {volume} {86}},\ \bibinfo
  {pages} {121} (\bibinfo {year} {2013})}\BibitemShut {NoStop}%
\bibitem [{\citenamefont {Garcia}\ \emph {et~al.}(2015)\citenamefont {Garcia},
  \citenamefont {Hannezo}, \citenamefont {Elgeti}, \citenamefont {Joanny},
  \citenamefont {Silberzan},\ and\ \citenamefont {Gov}}]{garcia2015physics}%
  \BibitemOpen
  \bibfield  {author} {\bibinfo {author} {\bibfnamefont {S.}~\bibnamefont
  {Garcia}}, \bibinfo {author} {\bibfnamefont {E.}~\bibnamefont {Hannezo}},
  \bibinfo {author} {\bibfnamefont {J.}~\bibnamefont {Elgeti}}, \bibinfo
  {author} {\bibfnamefont {J.-F.}\ \bibnamefont {Joanny}}, \bibinfo {author}
  {\bibfnamefont {P.}~\bibnamefont {Silberzan}},\ and\ \bibinfo {author}
  {\bibfnamefont {N.~S.}\ \bibnamefont {Gov}},\ }\bibfield  {title} {\bibinfo
  {title} {Physics of active jamming during collective cellular motion in a
  monolayer},\ }\href {https://doi.org/10.1073/pnas.1510973112} {\bibfield
  {journal} {\bibinfo  {journal} {Proc. Natl. Acad. Sci. U.S.A.}\ }\textbf
  {\bibinfo {volume} {112}},\ \bibinfo {pages} {15314} (\bibinfo {year}
  {2015})}\BibitemShut {NoStop}%
\bibitem [{\citenamefont {Camley}\ and\ \citenamefont
  {Rappel}(2017)}]{camley2017physical}%
  \BibitemOpen
  \bibfield  {author} {\bibinfo {author} {\bibfnamefont {B.~A.}\ \bibnamefont
  {Camley}}\ and\ \bibinfo {author} {\bibfnamefont {W.-J.}\ \bibnamefont
  {Rappel}},\ }\bibfield  {title} {\bibinfo {title} {Physical models of
  collective cell motility: from cell to tissue},\ }\href
  {https://doi.org/10.1088/1361-6463/aa56fe} {\bibfield  {journal} {\bibinfo
  {journal} {J. Phys. D}\ }\textbf {\bibinfo {volume} {50}},\ \bibinfo {pages}
  {113002} (\bibinfo {year} {2017})}\BibitemShut {NoStop}%
\bibitem [{\citenamefont {Lin}\ \emph {et~al.}(2017)\citenamefont {Lin},
  \citenamefont {Li}, \citenamefont {Xu},\ and\ \citenamefont
  {Feng}}]{lin2017collective}%
  \BibitemOpen
  \bibfield  {author} {\bibinfo {author} {\bibfnamefont {S.-Z.}\ \bibnamefont
  {Lin}}, \bibinfo {author} {\bibfnamefont {B.}~\bibnamefont {Li}}, \bibinfo
  {author} {\bibfnamefont {G.-K.}\ \bibnamefont {Xu}},\ and\ \bibinfo {author}
  {\bibfnamefont {X.-Q.}\ \bibnamefont {Feng}},\ }\bibfield  {title} {\bibinfo
  {title} {Collective dynamics of cancer cells confined in a confluent
  monolayer of normal cells},\ }\href
  {https://doi.org/10.1016/j.jbiomech.2016.12.035} {\bibfield  {journal}
  {\bibinfo  {journal} {J. Biomech.}\ }\textbf {\bibinfo {volume} {52}},\
  \bibinfo {pages} {140} (\bibinfo {year} {2017})}\BibitemShut {NoStop}%
\bibitem [{\citenamefont {Ladoux}\ and\ \citenamefont
  {M{\`e}ge}(2017)}]{ladoux2017mechanobiology}%
  \BibitemOpen
  \bibfield  {author} {\bibinfo {author} {\bibfnamefont {B.}~\bibnamefont
  {Ladoux}}\ and\ \bibinfo {author} {\bibfnamefont {R.-M.}\ \bibnamefont
  {M{\`e}ge}},\ }\bibfield  {title} {\bibinfo {title} {Mechanobiology of
  collective cell behaviours},\ }\href
  {https://doi.org/doi:10.1038/nrm.2017.98} {\bibfield  {journal} {\bibinfo
  {journal} {Nat. Rev. Mol.}\ }\textbf {\bibinfo {volume} {18}},\ \bibinfo
  {pages} {743} (\bibinfo {year} {2017})}\BibitemShut {NoStop}%
\bibitem [{\citenamefont {Stuelten}\ \emph {et~al.}(2018)\citenamefont
  {Stuelten}, \citenamefont {Parent},\ and\ \citenamefont
  {Montell}}]{stuelten2018cell}%
  \BibitemOpen
  \bibfield  {author} {\bibinfo {author} {\bibfnamefont {C.~H.}\ \bibnamefont
  {Stuelten}}, \bibinfo {author} {\bibfnamefont {C.~A.}\ \bibnamefont
  {Parent}},\ and\ \bibinfo {author} {\bibfnamefont {D.~J.}\ \bibnamefont
  {Montell}},\ }\bibfield  {title} {\bibinfo {title} {Cell motility in cancer
  invasion and metastasis: insights from simple model organisms},\ }\href
  {https://doi.org/10.1038/nrc.2018.15} {\bibfield  {journal} {\bibinfo
  {journal} {Nat. Rev. Cancer}\ }\textbf {\bibinfo {volume} {18}},\ \bibinfo
  {pages} {296} (\bibinfo {year} {2018})}\BibitemShut {NoStop}%
\bibitem [{\citenamefont {Hamby}(2019)}]{hamby2019connecting}%
  \BibitemOpen
  \bibfield  {author} {\bibinfo {author} {\bibfnamefont {A.~E.}\ \bibnamefont
  {Hamby}},\ }\href {http://hdl.handle.net/10150/634367} {\bibinfo {title}
  {Connecting the biophysics of active matter to collective migration}}
  (\bibinfo {year} {2019})\BibitemShut {NoStop}%
\bibitem [{\citenamefont {Spatarelu}\ \emph {et~al.}(2019)\citenamefont
  {Spatarelu}, \citenamefont {Zhang}, \citenamefont {Nguyen}, \citenamefont
  {Han}, \citenamefont {Liu}, \citenamefont {Guo}, \citenamefont {Notbohm},
  \citenamefont {Fan}, \citenamefont {Liu},\ and\ \citenamefont
  {Chen}}]{spatarelu2019biomechanics}%
  \BibitemOpen
  \bibfield  {author} {\bibinfo {author} {\bibfnamefont {C.-P.}\ \bibnamefont
  {Spatarelu}}, \bibinfo {author} {\bibfnamefont {H.}~\bibnamefont {Zhang}},
  \bibinfo {author} {\bibfnamefont {D.~T.}\ \bibnamefont {Nguyen}}, \bibinfo
  {author} {\bibfnamefont {X.}~\bibnamefont {Han}}, \bibinfo {author}
  {\bibfnamefont {R.}~\bibnamefont {Liu}}, \bibinfo {author} {\bibfnamefont
  {Q.}~\bibnamefont {Guo}}, \bibinfo {author} {\bibfnamefont {J.}~\bibnamefont
  {Notbohm}}, \bibinfo {author} {\bibfnamefont {J.}~\bibnamefont {Fan}},
  \bibinfo {author} {\bibfnamefont {L.}~\bibnamefont {Liu}},\ and\ \bibinfo
  {author} {\bibfnamefont {Z.}~\bibnamefont {Chen}},\ }\bibfield  {title}
  {\bibinfo {title} {Biomechanics of collective cell migration in cancer
  progression: Experimental and computational methods},\ }\href
  {https://doi.org/10.1021/acsbiomaterials.8b01428} {\bibfield  {journal}
  {\bibinfo  {journal} {ACS Biomater. Sci. Eng.}\ }\textbf {\bibinfo {volume}
  {5}},\ \bibinfo {pages} {3766} (\bibinfo {year} {2019})}\BibitemShut
  {NoStop}%
\bibitem [{\citenamefont {Volfson}\ \emph {et~al.}(2008)\citenamefont
  {Volfson}, \citenamefont {Cookson}, \citenamefont {Hasty},\ and\
  \citenamefont {Tsimring}}]{volfson2008biomechanical}%
  \BibitemOpen
  \bibfield  {author} {\bibinfo {author} {\bibfnamefont {D.}~\bibnamefont
  {Volfson}}, \bibinfo {author} {\bibfnamefont {S.}~\bibnamefont {Cookson}},
  \bibinfo {author} {\bibfnamefont {J.}~\bibnamefont {Hasty}},\ and\ \bibinfo
  {author} {\bibfnamefont {L.~S.}\ \bibnamefont {Tsimring}},\ }\bibfield
  {title} {\bibinfo {title} {Biomechanical ordering of dense cell
  populations},\ }\href {https://doi.org/10.1073/pnas.0706805105} {\bibfield
  {journal} {\bibinfo  {journal} {Proc. Natl. Acad. Sci. U.S.A}\ }\textbf
  {\bibinfo {volume} {105}},\ \bibinfo {pages} {15346} (\bibinfo {year}
  {2008})}\BibitemShut {NoStop}%
\bibitem [{\citenamefont {Bove}\ \emph {et~al.}(2017)\citenamefont {Bove},
  \citenamefont {Gradeci}, \citenamefont {Fujita}, \citenamefont {Banerjee},
  \citenamefont {Charras},\ and\ \citenamefont {Lowe}}]{bove2017local}%
  \BibitemOpen
  \bibfield  {author} {\bibinfo {author} {\bibfnamefont {A.}~\bibnamefont
  {Bove}}, \bibinfo {author} {\bibfnamefont {D.}~\bibnamefont {Gradeci}},
  \bibinfo {author} {\bibfnamefont {Y.}~\bibnamefont {Fujita}}, \bibinfo
  {author} {\bibfnamefont {S.}~\bibnamefont {Banerjee}}, \bibinfo {author}
  {\bibfnamefont {G.}~\bibnamefont {Charras}},\ and\ \bibinfo {author}
  {\bibfnamefont {A.~R.}\ \bibnamefont {Lowe}},\ }\bibfield  {title} {\bibinfo
  {title} {Local cellular neighborhood controls proliferation in cell
  competition},\ }\href {https://doi.org/10.1091/mbc.e17-06-0368} {\bibfield
  {journal} {\bibinfo  {journal} {Mol. Biol. Cell}\ }\textbf {\bibinfo {volume}
  {28}},\ \bibinfo {pages} {3215} (\bibinfo {year} {2017})}\BibitemShut
  {NoStop}%
\bibitem [{\citenamefont {Puliafito}\ \emph {et~al.}(2012)\citenamefont
  {Puliafito}, \citenamefont {Hufnagel}, \citenamefont {Neveu}, \citenamefont
  {Streichan}, \citenamefont {Sigal}, \citenamefont {Fygenson},\ and\
  \citenamefont {Shraiman}}]{puliafito2012collective}%
  \BibitemOpen
  \bibfield  {author} {\bibinfo {author} {\bibfnamefont {A.}~\bibnamefont
  {Puliafito}}, \bibinfo {author} {\bibfnamefont {L.}~\bibnamefont {Hufnagel}},
  \bibinfo {author} {\bibfnamefont {P.}~\bibnamefont {Neveu}}, \bibinfo
  {author} {\bibfnamefont {S.}~\bibnamefont {Streichan}}, \bibinfo {author}
  {\bibfnamefont {A.}~\bibnamefont {Sigal}}, \bibinfo {author} {\bibfnamefont
  {D.~K.}\ \bibnamefont {Fygenson}},\ and\ \bibinfo {author} {\bibfnamefont
  {B.~I.}\ \bibnamefont {Shraiman}},\ }\bibfield  {title} {\bibinfo {title}
  {Collective and single cell behavior in epithelial contact inhibition},\
  }\href {https://doi.org/10.1073/pnas.1007809109} {\bibfield  {journal}
  {\bibinfo  {journal} {Proc. Natl. Acad. Sci. U.S.A.}\ }\textbf {\bibinfo
  {volume} {109}},\ \bibinfo {pages} {739} (\bibinfo {year}
  {2012})}\BibitemShut {NoStop}%
\bibitem [{\citenamefont {Doostmohammadi}\ \emph {et~al.}(2015)\citenamefont
  {Doostmohammadi}, \citenamefont {Thampi}, \citenamefont {Saw}, \citenamefont
  {Lim}, \citenamefont {Ladoux},\ and\ \citenamefont
  {Yeomans}}]{doostmohammadi2015celebrating}%
  \BibitemOpen
  \bibfield  {author} {\bibinfo {author} {\bibfnamefont {A.}~\bibnamefont
  {Doostmohammadi}}, \bibinfo {author} {\bibfnamefont {S.~P.}\ \bibnamefont
  {Thampi}}, \bibinfo {author} {\bibfnamefont {T.~B.}\ \bibnamefont {Saw}},
  \bibinfo {author} {\bibfnamefont {C.~T.}\ \bibnamefont {Lim}}, \bibinfo
  {author} {\bibfnamefont {B.}~\bibnamefont {Ladoux}},\ and\ \bibinfo {author}
  {\bibfnamefont {J.~M.}\ \bibnamefont {Yeomans}},\ }\bibfield  {title}
  {\bibinfo {title} {Celebrating soft matter's 10th anniversary: Cell division:
  a source of active stress in cellular monolayers},\ }\href
  {https://doi.org/10.1039/c5sm01382h} {\bibfield  {journal} {\bibinfo
  {journal} {Soft Matter}\ }\textbf {\bibinfo {volume} {11}},\ \bibinfo {pages}
  {7328} (\bibinfo {year} {2015})}\BibitemShut {NoStop}%
\bibitem [{\citenamefont {Heinrich}\ \emph {et~al.}(2020)\citenamefont
  {Heinrich}, \citenamefont {LaChance}, \citenamefont {Zajdel}, \citenamefont
  {Alert}, \citenamefont {Kosmrlj},\ and\ \citenamefont
  {Cohen}}]{heinrich2020size}%
  \BibitemOpen
  \bibfield  {author} {\bibinfo {author} {\bibfnamefont {M.~A.}\ \bibnamefont
  {Heinrich}}, \bibinfo {author} {\bibfnamefont {J.}~\bibnamefont {LaChance}},
  \bibinfo {author} {\bibfnamefont {T.~J.}\ \bibnamefont {Zajdel}}, \bibinfo
  {author} {\bibfnamefont {R.}~\bibnamefont {Alert}}, \bibinfo {author}
  {\bibfnamefont {A.}~\bibnamefont {Kosmrlj}},\ and\ \bibinfo {author}
  {\bibfnamefont {D.~J.}\ \bibnamefont {Cohen}},\ }\bibfield  {title} {\bibinfo
  {title} {Size-dependent patterns of cell proliferation and migration in
  freely-expanding epithelia},\ }\bibfield  {journal} {\bibinfo  {journal}
  {Elife}\ }\textbf {\bibinfo {volume} {9}},\ \href
  {https://doi.org/10.7554/eLife.58945} {10.7554/eLife.58945} (\bibinfo {year}
  {2020})\BibitemShut {NoStop}%
\bibitem [{\citenamefont {Trepat}\ \emph {et~al.}(2009)\citenamefont {Trepat},
  \citenamefont {Wasserman}, \citenamefont {Angelini}, \citenamefont {Millet},
  \citenamefont {Weitz}, \citenamefont {Butler},\ and\ \citenamefont
  {Fredberg}}]{trepat2009physical}%
  \BibitemOpen
  \bibfield  {author} {\bibinfo {author} {\bibfnamefont {X.}~\bibnamefont
  {Trepat}}, \bibinfo {author} {\bibfnamefont {M.~R.}\ \bibnamefont
  {Wasserman}}, \bibinfo {author} {\bibfnamefont {T.~E.}\ \bibnamefont
  {Angelini}}, \bibinfo {author} {\bibfnamefont {E.}~\bibnamefont {Millet}},
  \bibinfo {author} {\bibfnamefont {D.~A.}\ \bibnamefont {Weitz}}, \bibinfo
  {author} {\bibfnamefont {J.~P.}\ \bibnamefont {Butler}},\ and\ \bibinfo
  {author} {\bibfnamefont {J.~J.}\ \bibnamefont {Fredberg}},\ }\bibfield
  {title} {\bibinfo {title} {Physical forces during collective cell
  migration},\ }\href {https://doi.org/10.1038/NPHYS1269} {\bibfield  {journal}
  {\bibinfo  {journal} {Nat. Phys.}\ }\textbf {\bibinfo {volume} {5}},\
  \bibinfo {pages} {426} (\bibinfo {year} {2009})}\BibitemShut {NoStop}%
\bibitem [{\citenamefont {Luo}\ \emph {et~al.}(2013)\citenamefont {Luo},
  \citenamefont {Mohan}, \citenamefont {Iglesias},\ and\ \citenamefont
  {Robinson}}]{luo2013molecular}%
  \BibitemOpen
  \bibfield  {author} {\bibinfo {author} {\bibfnamefont {T.}~\bibnamefont
  {Luo}}, \bibinfo {author} {\bibfnamefont {K.}~\bibnamefont {Mohan}}, \bibinfo
  {author} {\bibfnamefont {P.~A.}\ \bibnamefont {Iglesias}},\ and\ \bibinfo
  {author} {\bibfnamefont {D.~N.}\ \bibnamefont {Robinson}},\ }\bibfield
  {title} {\bibinfo {title} {Molecular mechanisms of cellular mechanosensing},\
  }\href {https://doi.org/10.1038/NMAT3772} {\bibfield  {journal} {\bibinfo
  {journal} {Nat. Mater.}\ }\textbf {\bibinfo {volume} {12}},\ \bibinfo {pages}
  {1064} (\bibinfo {year} {2013})}\BibitemShut {NoStop}%
\bibitem [{\citenamefont {Mao}\ and\ \citenamefont {Baum}(2015)}]{mao2015tug}%
  \BibitemOpen
  \bibfield  {author} {\bibinfo {author} {\bibfnamefont {Y.}~\bibnamefont
  {Mao}}\ and\ \bibinfo {author} {\bibfnamefont {B.}~\bibnamefont {Baum}},\
  }\bibfield  {title} {\bibinfo {title} {Tug of war—the influence of opposing
  physical forces on epithelial cell morphology},\ }\href
  {https://doi.org/10.1016/j.ydbio.2014.12.030} {\bibfield  {journal} {\bibinfo
   {journal} {Dev. Biol.}\ }\textbf {\bibinfo {volume} {401}},\ \bibinfo
  {pages} {92} (\bibinfo {year} {2015})}\BibitemShut {NoStop}%
\bibitem [{\citenamefont {Persat}\ \emph {et~al.}(2015)\citenamefont {Persat},
  \citenamefont {Nadell}, \citenamefont {Kim}, \citenamefont {Ingremeau},
  \citenamefont {Siryaporn}, \citenamefont {Drescher}, \citenamefont
  {Wingreen}, \citenamefont {Bassler}, \citenamefont {Gitai},\ and\
  \citenamefont {Stone}}]{persat2015mechanical}%
  \BibitemOpen
  \bibfield  {author} {\bibinfo {author} {\bibfnamefont {A.}~\bibnamefont
  {Persat}}, \bibinfo {author} {\bibfnamefont {C.~D.}\ \bibnamefont {Nadell}},
  \bibinfo {author} {\bibfnamefont {M.~K.}\ \bibnamefont {Kim}}, \bibinfo
  {author} {\bibfnamefont {F.}~\bibnamefont {Ingremeau}}, \bibinfo {author}
  {\bibfnamefont {A.}~\bibnamefont {Siryaporn}}, \bibinfo {author}
  {\bibfnamefont {K.}~\bibnamefont {Drescher}}, \bibinfo {author}
  {\bibfnamefont {N.~S.}\ \bibnamefont {Wingreen}}, \bibinfo {author}
  {\bibfnamefont {B.~L.}\ \bibnamefont {Bassler}}, \bibinfo {author}
  {\bibfnamefont {Z.}~\bibnamefont {Gitai}},\ and\ \bibinfo {author}
  {\bibfnamefont {H.~A.}\ \bibnamefont {Stone}},\ }\bibfield  {title} {\bibinfo
  {title} {The mechanical world of bacteria},\ }\href
  {https://doi.org/10.1016/j.cell.2015.05.005} {\bibfield  {journal} {\bibinfo
  {journal} {Cell}\ }\textbf {\bibinfo {volume} {161}},\ \bibinfo {pages} {988}
  (\bibinfo {year} {2015})}\BibitemShut {NoStop}%
\bibitem [{\citenamefont {Delarue}\ \emph {et~al.}(2016)\citenamefont
  {Delarue}, \citenamefont {Hartung}, \citenamefont {Schreck}, \citenamefont
  {Gniewek}, \citenamefont {Hu}, \citenamefont {Herminghaus},\ and\
  \citenamefont {Hallatschek}}]{delarue2016self}%
  \BibitemOpen
  \bibfield  {author} {\bibinfo {author} {\bibfnamefont {M.}~\bibnamefont
  {Delarue}}, \bibinfo {author} {\bibfnamefont {J.}~\bibnamefont {Hartung}},
  \bibinfo {author} {\bibfnamefont {C.}~\bibnamefont {Schreck}}, \bibinfo
  {author} {\bibfnamefont {P.}~\bibnamefont {Gniewek}}, \bibinfo {author}
  {\bibfnamefont {L.}~\bibnamefont {Hu}}, \bibinfo {author} {\bibfnamefont
  {S.}~\bibnamefont {Herminghaus}},\ and\ \bibinfo {author} {\bibfnamefont
  {O.}~\bibnamefont {Hallatschek}},\ }\bibfield  {title} {\bibinfo {title}
  {Self-driven jamming in growing microbial populations},\ }\href
  {https://doi.org/10.1038/NPHYS3741} {\bibfield  {journal} {\bibinfo
  {journal} {Nat. Phys.}\ }\textbf {\bibinfo {volume} {12}},\ \bibinfo {pages}
  {762} (\bibinfo {year} {2016})}\BibitemShut {NoStop}%
\bibitem [{\citenamefont {Wang}(2017)}]{wang2017review}%
  \BibitemOpen
  \bibfield  {author} {\bibinfo {author} {\bibfnamefont {N.}~\bibnamefont
  {Wang}},\ }\bibfield  {title} {\bibinfo {title} {Review of cellular
  mechanotransduction},\ }\href {https://doi.org/10.1088/1361-6463/aa6e18}
  {\bibfield  {journal} {\bibinfo  {journal} {J. Phys. D}\ }\textbf {\bibinfo
  {volume} {50}},\ \bibinfo {pages} {233002} (\bibinfo {year}
  {2017})}\BibitemShut {NoStop}%
\bibitem [{\citenamefont {Irvine}\ and\ \citenamefont
  {Shraiman}(2017)}]{irvine2017mechanical}%
  \BibitemOpen
  \bibfield  {author} {\bibinfo {author} {\bibfnamefont {K.~D.}\ \bibnamefont
  {Irvine}}\ and\ \bibinfo {author} {\bibfnamefont {B.~I.}\ \bibnamefont
  {Shraiman}},\ }\bibfield  {title} {\bibinfo {title} {Mechanical control of
  growth: ideas, facts and challenges},\ }\href
  {https://doi.org/10.1242/dev.151902} {\bibfield  {journal} {\bibinfo
  {journal} {Development}\ }\textbf {\bibinfo {volume} {144}},\ \bibinfo
  {pages} {4238} (\bibinfo {year} {2017})}\BibitemShut {NoStop}%
\bibitem [{\citenamefont {Chen}\ \emph {et~al.}(2019)\citenamefont {Chen},
  \citenamefont {Li},\ and\ \citenamefont {Ju}}]{chen2019tensile}%
  \BibitemOpen
  \bibfield  {author} {\bibinfo {author} {\bibfnamefont {Y.}~\bibnamefont
  {Chen}}, \bibinfo {author} {\bibfnamefont {Z.}~\bibnamefont {Li}},\ and\
  \bibinfo {author} {\bibfnamefont {L.~A.}\ \bibnamefont {Ju}},\ }\bibfield
  {title} {\bibinfo {title} {Tensile and compressive force regulation on cell
  mechanosensing},\ }\href {https://doi.org/10.1007/s12551-019-00536-z}
  {\bibfield  {journal} {\bibinfo  {journal} {Biophys. Rev.}\ ,\ \bibinfo
  {pages} {1}} (\bibinfo {year} {2019})}\BibitemShut {NoStop}%
\bibitem [{\citenamefont {Hamouche}\ \emph {et~al.}(2017)\citenamefont
  {Hamouche}, \citenamefont {Laalami}, \citenamefont {Daerr}, \citenamefont
  {Song}, \citenamefont {Holland}, \citenamefont {S{\'e}ror}, \citenamefont
  {Hamze},\ and\ \citenamefont {Putzer}}]{hamouche2017bacillus}%
  \BibitemOpen
  \bibfield  {author} {\bibinfo {author} {\bibfnamefont {L.}~\bibnamefont
  {Hamouche}}, \bibinfo {author} {\bibfnamefont {S.}~\bibnamefont {Laalami}},
  \bibinfo {author} {\bibfnamefont {A.}~\bibnamefont {Daerr}}, \bibinfo
  {author} {\bibfnamefont {S.}~\bibnamefont {Song}}, \bibinfo {author}
  {\bibfnamefont {I.~B.}\ \bibnamefont {Holland}}, \bibinfo {author}
  {\bibfnamefont {S.~J.}\ \bibnamefont {S{\'e}ror}}, \bibinfo {author}
  {\bibfnamefont {K.}~\bibnamefont {Hamze}},\ and\ \bibinfo {author}
  {\bibfnamefont {H.}~\bibnamefont {Putzer}},\ }\bibfield  {title} {\bibinfo
  {title} {Bacillus subtilis swarmer cells lead the swarm, multiply, and
  generate a trail of quiescent descendants},\ }\href
  {https://doi.org/10.1128/mBio.02102-16} {\bibfield  {journal} {\bibinfo
  {journal} {Mbio}\ }\textbf {\bibinfo {volume} {8}},\ \bibinfo {pages}
  {e02102} (\bibinfo {year} {2017})}\BibitemShut {NoStop}%
\bibitem [{\citenamefont {Yabunaka}\ and\ \citenamefont
  {Marcq}(2017)}]{yabunaka2017cell}%
  \BibitemOpen
  \bibfield  {author} {\bibinfo {author} {\bibfnamefont {S.}~\bibnamefont
  {Yabunaka}}\ and\ \bibinfo {author} {\bibfnamefont {P.}~\bibnamefont
  {Marcq}},\ }\bibfield  {title} {\bibinfo {title} {Cell growth, division, and
  death in cohesive tissues: A thermodynamic approach},\ }\href
  {https://doi.org/10.1103/PhysRevE.96.022406} {\bibfield  {journal} {\bibinfo
  {journal} {Phys. Rev. E}\ }\textbf {\bibinfo {volume} {96}},\ \bibinfo
  {pages} {022406} (\bibinfo {year} {2017})}\BibitemShut {NoStop}%
\bibitem [{\citenamefont {Malmi-Kakkada}\ \emph {et~al.}(2018)\citenamefont
  {Malmi-Kakkada}, \citenamefont {Li}, \citenamefont {Samanta}, \citenamefont
  {Sinha},\ and\ \citenamefont {Thirumalai}}]{malmi2018cell}%
  \BibitemOpen
  \bibfield  {author} {\bibinfo {author} {\bibfnamefont {A.~N.}\ \bibnamefont
  {Malmi-Kakkada}}, \bibinfo {author} {\bibfnamefont {X.}~\bibnamefont {Li}},
  \bibinfo {author} {\bibfnamefont {H.~S.}\ \bibnamefont {Samanta}}, \bibinfo
  {author} {\bibfnamefont {S.}~\bibnamefont {Sinha}},\ and\ \bibinfo {author}
  {\bibfnamefont {D.}~\bibnamefont {Thirumalai}},\ }\bibfield  {title}
  {\bibinfo {title} {Cell growth rate dictates the onset of glass to fluidlike
  transition and long time superdiffusion in an evolving cell colony},\ }\href
  {https://doi.org/10.1103/PhysRevX.8.021025} {\bibfield  {journal} {\bibinfo
  {journal} {Phys. Rev. X}\ }\textbf {\bibinfo {volume} {8}},\ \bibinfo {pages}
  {021025} (\bibinfo {year} {2018})}\BibitemShut {NoStop}%
\bibitem [{\citenamefont {Srinivasan}\ \emph {et~al.}(2019)\citenamefont
  {Srinivasan}, \citenamefont {Kaplan},\ and\ \citenamefont
  {Mahadevan}}]{srinivasan2019multiphase}%
  \BibitemOpen
  \bibfield  {author} {\bibinfo {author} {\bibfnamefont {S.}~\bibnamefont
  {Srinivasan}}, \bibinfo {author} {\bibfnamefont {C.~N.}\ \bibnamefont
  {Kaplan}},\ and\ \bibinfo {author} {\bibfnamefont {L.}~\bibnamefont
  {Mahadevan}},\ }\bibfield  {title} {\bibinfo {title} {A multiphase theory for
  spreading microbial swarms and films},\ }\href
  {https://doi.org/10.7554/eLife.42697} {\bibfield  {journal} {\bibinfo
  {journal} {Elife}\ }\textbf {\bibinfo {volume} {8}},\ \bibinfo {pages}
  {e42697} (\bibinfo {year} {2019})}\BibitemShut {NoStop}%
\bibitem [{\citenamefont {Ben~Amar}\ \emph {et~al.}(2019)\citenamefont
  {Ben~Amar}, \citenamefont {Nassoy},\ and\ \citenamefont
  {LeGoff}}]{ben2019physics}%
  \BibitemOpen
  \bibfield  {author} {\bibinfo {author} {\bibfnamefont {M.}~\bibnamefont
  {Ben~Amar}}, \bibinfo {author} {\bibfnamefont {P.}~\bibnamefont {Nassoy}},\
  and\ \bibinfo {author} {\bibfnamefont {L.}~\bibnamefont {LeGoff}},\
  }\bibfield  {title} {\bibinfo {title} {Physics of growing biological tissues:
  the complex cross-talk between cell activity, growth and resistance},\ }\href
  {https://doi.org/10.1098/rsta.2018.0070} {\bibfield  {journal} {\bibinfo
  {journal} {Philos. Trans. R. Soc. A}\ }\textbf {\bibinfo {volume} {377}},\
  \bibinfo {pages} {20180070} (\bibinfo {year} {2019})}\BibitemShut {NoStop}%
\bibitem [{\citenamefont {Mather}\ \emph {et~al.}(2010)\citenamefont {Mather},
  \citenamefont {Mondrag{\'o}n-Palomino}, \citenamefont {Danino}, \citenamefont
  {Hasty},\ and\ \citenamefont {Tsimring}}]{mather2010streaming}%
  \BibitemOpen
  \bibfield  {author} {\bibinfo {author} {\bibfnamefont {W.}~\bibnamefont
  {Mather}}, \bibinfo {author} {\bibfnamefont {O.}~\bibnamefont
  {Mondrag{\'o}n-Palomino}}, \bibinfo {author} {\bibfnamefont {T.}~\bibnamefont
  {Danino}}, \bibinfo {author} {\bibfnamefont {J.}~\bibnamefont {Hasty}},\ and\
  \bibinfo {author} {\bibfnamefont {L.~S.}\ \bibnamefont {Tsimring}},\
  }\bibfield  {title} {\bibinfo {title} {Streaming instability in growing cell
  populations},\ }\href {https://doi.org/10.1103/PhysRevLett.104.208101}
  {\bibfield  {journal} {\bibinfo  {journal} {Phys. Rev. Lett.}\ }\textbf
  {\bibinfo {volume} {104}},\ \bibinfo {pages} {208101} (\bibinfo {year}
  {2010})}\BibitemShut {NoStop}%
\bibitem [{\citenamefont {Dell’Arciprete}\ \emph {et~al.}(2018)\citenamefont
  {Dell’Arciprete}, \citenamefont {Blow}, \citenamefont {Brown},
  \citenamefont {Farrell}, \citenamefont {Lintuvuori}, \citenamefont {McVey},
  \citenamefont {Marenduzzo},\ and\ \citenamefont {Poon}}]{dell2018growing}%
  \BibitemOpen
  \bibfield  {author} {\bibinfo {author} {\bibfnamefont {D.}~\bibnamefont
  {Dell’Arciprete}}, \bibinfo {author} {\bibfnamefont {M.}~\bibnamefont
  {Blow}}, \bibinfo {author} {\bibfnamefont {A.}~\bibnamefont {Brown}},
  \bibinfo {author} {\bibfnamefont {F.}~\bibnamefont {Farrell}}, \bibinfo
  {author} {\bibfnamefont {J.~S.}\ \bibnamefont {Lintuvuori}}, \bibinfo
  {author} {\bibfnamefont {A.}~\bibnamefont {McVey}}, \bibinfo {author}
  {\bibfnamefont {D.}~\bibnamefont {Marenduzzo}},\ and\ \bibinfo {author}
  {\bibfnamefont {W.~C.}\ \bibnamefont {Poon}},\ }\bibfield  {title} {\bibinfo
  {title} {A growing bacterial colony in two dimensions as an active nematic},\
  }\href {https://doi.org/10.1038/s41467-018-06370-3} {\bibfield  {journal}
  {\bibinfo  {journal} {Nat. Commun.}\ }\textbf {\bibinfo {volume} {9}},\
  \bibinfo {pages} {1} (\bibinfo {year} {2018})}\BibitemShut {NoStop}%
\bibitem [{\citenamefont {Marinkovic}\ \emph {et~al.}(2019)\citenamefont
  {Marinkovic}, \citenamefont {Vulin}, \citenamefont {Acman}, \citenamefont
  {Song}, \citenamefont {Di~Meglio}, \citenamefont {Lindner},\ and\
  \citenamefont {Hersen}}]{marinkovic2019microfluidic}%
  \BibitemOpen
  \bibfield  {author} {\bibinfo {author} {\bibfnamefont {Z.~S.}\ \bibnamefont
  {Marinkovic}}, \bibinfo {author} {\bibfnamefont {C.}~\bibnamefont {Vulin}},
  \bibinfo {author} {\bibfnamefont {M.}~\bibnamefont {Acman}}, \bibinfo
  {author} {\bibfnamefont {X.}~\bibnamefont {Song}}, \bibinfo {author}
  {\bibfnamefont {J.-M.}\ \bibnamefont {Di~Meglio}}, \bibinfo {author}
  {\bibfnamefont {A.~B.}\ \bibnamefont {Lindner}},\ and\ \bibinfo {author}
  {\bibfnamefont {P.}~\bibnamefont {Hersen}},\ }\bibfield  {title} {\bibinfo
  {title} {A microfluidic device for inferring metabolic landscapes in yeast
  monolayer colonies},\ }\bibfield  {journal} {\bibinfo  {journal} {Elife}\
  }\textbf {\bibinfo {volume} {8}},\ \href
  {https://doi.org/10.7554/eLife.47951} {10.7554/eLife.47951} (\bibinfo {year}
  {2019})\BibitemShut {NoStop}%
\bibitem [{\citenamefont {Xi}\ \emph {et~al.}(2017)\citenamefont {Xi},
  \citenamefont {Sonam}, \citenamefont {Saw}, \citenamefont {Ladoux},\ and\
  \citenamefont {Lim}}]{xi2017emergent}%
  \BibitemOpen
  \bibfield  {author} {\bibinfo {author} {\bibfnamefont {W.}~\bibnamefont
  {Xi}}, \bibinfo {author} {\bibfnamefont {S.}~\bibnamefont {Sonam}}, \bibinfo
  {author} {\bibfnamefont {T.~B.}\ \bibnamefont {Saw}}, \bibinfo {author}
  {\bibfnamefont {B.}~\bibnamefont {Ladoux}},\ and\ \bibinfo {author}
  {\bibfnamefont {C.~T.}\ \bibnamefont {Lim}},\ }\bibfield  {title} {\bibinfo
  {title} {Emergent patterns of collective cell migration under tubular
  confinement},\ }\href {https://doi.org/10.1038/s41467-017-01390-x} {\bibfield
   {journal} {\bibinfo  {journal} {Nat. Commun.}\ }\textbf {\bibinfo {volume}
  {8}},\ \bibinfo {pages} {1} (\bibinfo {year} {2017})}\BibitemShut {NoStop}%
\bibitem [{\citenamefont {Huergo}\ \emph {et~al.}(2011)\citenamefont {Huergo},
  \citenamefont {Pasquale}, \citenamefont {Gonzalez}, \citenamefont {Bolzan},\
  and\ \citenamefont {Arvia}}]{huergo2011dynamics}%
  \BibitemOpen
  \bibfield  {author} {\bibinfo {author} {\bibfnamefont {M.~A.~C.}\
  \bibnamefont {Huergo}}, \bibinfo {author} {\bibfnamefont {M.~A.}\
  \bibnamefont {Pasquale}}, \bibinfo {author} {\bibfnamefont {P.~H.}\
  \bibnamefont {Gonzalez}}, \bibinfo {author} {\bibfnamefont {A.~E.}\
  \bibnamefont {Bolzan}},\ and\ \bibinfo {author} {\bibfnamefont {A.~J.}\
  \bibnamefont {Arvia}},\ }\bibfield  {title} {\bibinfo {title} {Dynamics and
  morphology characteristics of cell colonies with radially spreading growth
  fronts},\ }\href {https://doi.org/10.1103/PhysRevE.84.021917} {\bibfield
  {journal} {\bibinfo  {journal} {Phys. Rev. E}\ }\textbf {\bibinfo {volume}
  {84}},\ \bibinfo {pages} {021917} (\bibinfo {year} {2011})}\BibitemShut
  {NoStop}%
\bibitem [{\citenamefont {Gauquelin}\ \emph {et~al.}(2019)\citenamefont
  {Gauquelin}, \citenamefont {Tlili}, \citenamefont {Gay}, \citenamefont
  {Peyret}, \citenamefont {M{\`{e}}ge}, \citenamefont {Fardin},\ and\
  \citenamefont {Ladoux}}]{Gauquelin2019}%
  \BibitemOpen
  \bibfield  {author} {\bibinfo {author} {\bibfnamefont {E.}~\bibnamefont
  {Gauquelin}}, \bibinfo {author} {\bibfnamefont {S.}~\bibnamefont {Tlili}},
  \bibinfo {author} {\bibfnamefont {C.}~\bibnamefont {Gay}}, \bibinfo {author}
  {\bibfnamefont {G.}~\bibnamefont {Peyret}}, \bibinfo {author} {\bibfnamefont
  {R.-M.}\ \bibnamefont {M{\`{e}}ge}}, \bibinfo {author} {\bibfnamefont
  {M.~A.}\ \bibnamefont {Fardin}},\ and\ \bibinfo {author} {\bibfnamefont
  {B.}~\bibnamefont {Ladoux}},\ }\bibfield  {title} {\bibinfo {title}
  {Influence of proliferation on the motions of epithelial monolayers invading
  adherent strips},\ }\href {https://doi.org/10.1039/C9SM00105K} {\bibfield
  {journal} {\bibinfo  {journal} {Soft Matter}\ }\textbf {\bibinfo {volume}
  {15}},\ \bibinfo {pages} {2798} (\bibinfo {year} {2019})}\BibitemShut
  {NoStop}%
\bibitem [{\citenamefont {Wang}\ and\ \citenamefont
  {Levin}(2009)}]{wang2009metabolism}%
  \BibitemOpen
  \bibfield  {author} {\bibinfo {author} {\bibfnamefont {J.~D.}\ \bibnamefont
  {Wang}}\ and\ \bibinfo {author} {\bibfnamefont {P.~A.}\ \bibnamefont
  {Levin}},\ }\bibfield  {title} {\bibinfo {title} {Metabolism, cell growth and
  the bacterial cell cycle},\ }\href {https://doi.org/10.1038/nrmicro2202}
  {\bibfield  {journal} {\bibinfo  {journal} {Nat. Rev. Microbiol.}\ }\textbf
  {\bibinfo {volume} {7}},\ \bibinfo {pages} {822} (\bibinfo {year}
  {2009})}\BibitemShut {NoStop}%
\bibitem [{\citenamefont {Theveneau}\ and\ \citenamefont
  {Mayor}(2013)}]{theveneau2013collective}%
  \BibitemOpen
  \bibfield  {author} {\bibinfo {author} {\bibfnamefont {E.}~\bibnamefont
  {Theveneau}}\ and\ \bibinfo {author} {\bibfnamefont {R.}~\bibnamefont
  {Mayor}},\ }\bibfield  {title} {\bibinfo {title} {Collective cell migration
  of epithelial and mesenchymal cells},\ }\href
  {https://doi.org/10.1007/s00018-012-1251-7} {\bibfield  {journal} {\bibinfo
  {journal} {Cell. Mol. Life Sci.}\ }\textbf {\bibinfo {volume} {70}},\
  \bibinfo {pages} {3481} (\bibinfo {year} {2013})}\BibitemShut {NoStop}%
\bibitem [{\citenamefont {Benham-Pyle}\ \emph {et~al.}(2015)\citenamefont
  {Benham-Pyle}, \citenamefont {Pruitt},\ and\ \citenamefont
  {Nelson}}]{benham2015mechanical}%
  \BibitemOpen
  \bibfield  {author} {\bibinfo {author} {\bibfnamefont {B.~W.}\ \bibnamefont
  {Benham-Pyle}}, \bibinfo {author} {\bibfnamefont {B.~L.}\ \bibnamefont
  {Pruitt}},\ and\ \bibinfo {author} {\bibfnamefont {W.~J.}\ \bibnamefont
  {Nelson}},\ }\bibfield  {title} {\bibinfo {title} {Mechanical strain induces
  e-cadherin--dependent yap1 and $\beta$-catenin activation to drive cell cycle
  entry},\ }\href@noop {} {\bibfield  {journal} {\bibinfo  {journal} {Science}\
  }\textbf {\bibinfo {volume} {348}},\ \bibinfo {pages} {1024} (\bibinfo {year}
  {2015})}\BibitemShut {NoStop}%
\bibitem [{\citenamefont {Morgan}(2007)}]{morgan2007cell}%
  \BibitemOpen
  \bibfield  {author} {\bibinfo {author} {\bibfnamefont {D.~O.}\ \bibnamefont
  {Morgan}},\ }\href {https://doi.org/10.1093/icb/icm066} {\emph {\bibinfo
  {title} {The cell cycle: principles of control}}}\ (\bibinfo  {publisher}
  {New science press},\ \bibinfo {year} {2007})\BibitemShut {NoStop}%
\bibitem [{\citenamefont {Fiore}\ \emph {et~al.}(2018)\citenamefont {Fiore},
  \citenamefont {Ribeiro},\ and\ \citenamefont
  {Bruni-Cardoso}}]{fiore2018sleeping}%
  \BibitemOpen
  \bibfield  {author} {\bibinfo {author} {\bibfnamefont {A.~P. Z.~P.}\
  \bibnamefont {Fiore}}, \bibinfo {author} {\bibfnamefont {P.~d.~F.}\
  \bibnamefont {Ribeiro}},\ and\ \bibinfo {author} {\bibfnamefont
  {A.}~\bibnamefont {Bruni-Cardoso}},\ }\bibfield  {title} {\bibinfo {title}
  {Sleeping beauty and the microenvironment enchantment: Microenvironmental
  regulation of the proliferation-quiescence decision in normal tissues and in
  cancer development},\ }\href {https://doi.org/10.3389/fcell.2018.00059}
  {\bibfield  {journal} {\bibinfo  {journal} {Front. Cell Dev. Biol.}\ }\textbf
  {\bibinfo {volume} {6}},\ \bibinfo {pages} {59} (\bibinfo {year}
  {2018})}\BibitemShut {NoStop}%
\bibitem [{\citenamefont {Mathieu}\ and\ \citenamefont
  {Manneville}(2019)}]{mathieu2019intracellular}%
  \BibitemOpen
  \bibfield  {author} {\bibinfo {author} {\bibfnamefont {S.}~\bibnamefont
  {Mathieu}}\ and\ \bibinfo {author} {\bibfnamefont {J.-B.}\ \bibnamefont
  {Manneville}},\ }\bibfield  {title} {\bibinfo {title} {Intracellular
  mechanics: connecting rheology and mechanotransduction},\ }\href
  {https://doi.org/10.1016/j.ceb.2018.08.007} {\bibfield  {journal} {\bibinfo
  {journal} {Current opinion in cell biology}\ }\textbf {\bibinfo {volume}
  {56}},\ \bibinfo {pages} {34} (\bibinfo {year} {2019})}\BibitemShut {NoStop}%
\bibitem [{\citenamefont {Uroz}\ \emph
  {et~al.}(2018{\natexlab{a}})\citenamefont {Uroz}, \citenamefont {Wistorf},
  \citenamefont {Serra-Picamal}, \citenamefont {Conte}, \citenamefont
  {Sales-Pardo}, \citenamefont {Roca-Cusachs}, \citenamefont {Guimer{\`a}},\
  and\ \citenamefont {Trepat}}]{uroz2018regulation}%
  \BibitemOpen
  \bibfield  {author} {\bibinfo {author} {\bibfnamefont {M.}~\bibnamefont
  {Uroz}}, \bibinfo {author} {\bibfnamefont {S.}~\bibnamefont {Wistorf}},
  \bibinfo {author} {\bibfnamefont {X.}~\bibnamefont {Serra-Picamal}}, \bibinfo
  {author} {\bibfnamefont {V.}~\bibnamefont {Conte}}, \bibinfo {author}
  {\bibfnamefont {M.}~\bibnamefont {Sales-Pardo}}, \bibinfo {author}
  {\bibfnamefont {P.}~\bibnamefont {Roca-Cusachs}}, \bibinfo {author}
  {\bibfnamefont {R.}~\bibnamefont {Guimer{\`a}}},\ and\ \bibinfo {author}
  {\bibfnamefont {X.}~\bibnamefont {Trepat}},\ }\bibfield  {title} {\bibinfo
  {title} {Regulation of cell cycle progression by cell--cell and cell--matrix
  forces},\ }\href {https://doi.org/10.1038/s41556-018-0107-2} {\bibfield
  {journal} {\bibinfo  {journal} {Nat. Cell Biol.}\ }\textbf {\bibinfo {volume}
  {20}},\ \bibinfo {pages} {646} (\bibinfo {year}
  {2018}{\natexlab{a}})}\BibitemShut {NoStop}%
\bibitem [{\citenamefont {Otto}\ and\ \citenamefont
  {Sicinski}(2017)}]{otto2017cell}%
  \BibitemOpen
  \bibfield  {author} {\bibinfo {author} {\bibfnamefont {T.}~\bibnamefont
  {Otto}}\ and\ \bibinfo {author} {\bibfnamefont {P.}~\bibnamefont
  {Sicinski}},\ }\bibfield  {title} {\bibinfo {title} {Cell cycle proteins as
  promising targets in cancer therapy},\ }\href
  {https://doi.org/10.1038/nrc.2016.138} {\bibfield  {journal} {\bibinfo
  {journal} {Nature Reviews Cancer}\ }\textbf {\bibinfo {volume} {17}},\
  \bibinfo {pages} {93} (\bibinfo {year} {2017})}\BibitemShut {NoStop}%
\bibitem [{\citenamefont {Kent}\ and\ \citenamefont
  {Leone}(2019)}]{kent2019broken}%
  \BibitemOpen
  \bibfield  {author} {\bibinfo {author} {\bibfnamefont {L.~N.}\ \bibnamefont
  {Kent}}\ and\ \bibinfo {author} {\bibfnamefont {G.}~\bibnamefont {Leone}},\
  }\bibfield  {title} {\bibinfo {title} {The broken cycle: E2f dysfunction in
  cancer},\ }\href {https://doi.org/10.1038/s41568-019-0143-7} {\bibfield
  {journal} {\bibinfo  {journal} {Nature Reviews Cancer}\ }\textbf {\bibinfo
  {volume} {19}},\ \bibinfo {pages} {326} (\bibinfo {year} {2019})}\BibitemShut
  {NoStop}%
\bibitem [{\citenamefont {Devany}\ \emph {et~al.}(2021)\citenamefont {Devany},
  \citenamefont {Sussman}, \citenamefont {Yamamoto}, \citenamefont {Manning},\
  and\ \citenamefont {Gardel}}]{Devany2021}%
  \BibitemOpen
  \bibfield  {author} {\bibinfo {author} {\bibfnamefont {J.}~\bibnamefont
  {Devany}}, \bibinfo {author} {\bibfnamefont {D.~M.}\ \bibnamefont {Sussman}},
  \bibinfo {author} {\bibfnamefont {T.}~\bibnamefont {Yamamoto}}, \bibinfo
  {author} {\bibfnamefont {M.~L.}\ \bibnamefont {Manning}},\ and\ \bibinfo
  {author} {\bibfnamefont {M.~L.}\ \bibnamefont {Gardel}},\ }\bibfield  {title}
  {\bibinfo {title} {{Cell cycle–dependent active stress drives epithelia
  remodeling}},\ }\bibfield  {journal} {\bibinfo  {journal} {Proceedings of the
  National Academy of Sciences of the United States of America}\ }\textbf
  {\bibinfo {volume} {118}},\ \href {https://doi.org/10.1073/pnas.1917853118}
  {10.1073/pnas.1917853118} (\bibinfo {year} {2021})\BibitemShut {NoStop}%
\bibitem [{\citenamefont {Locke}\ \emph {et~al.}(2005)\citenamefont {Locke},
  \citenamefont {Millar},\ and\ \citenamefont {Turner}}]{LOCKE2005383}%
  \BibitemOpen
  \bibfield  {author} {\bibinfo {author} {\bibfnamefont {J.}~\bibnamefont
  {Locke}}, \bibinfo {author} {\bibfnamefont {A.}~\bibnamefont {Millar}},\ and\
  \bibinfo {author} {\bibfnamefont {M.}~\bibnamefont {Turner}},\ }\bibfield
  {title} {\bibinfo {title} {Modelling genetic networks with noisy and varied
  experimental data: the circadian clock in arabidopsis thaliana},\ }\href
  {https://doi.org/https://doi.org/10.1016/j.jtbi.2004.11.038} {\bibfield
  {journal} {\bibinfo  {journal} {J. Theor. Biol.}\ }\textbf {\bibinfo {volume}
  {234}},\ \bibinfo {pages} {383 } (\bibinfo {year} {2005})}\BibitemShut
  {NoStop}%
\bibitem [{\citenamefont {Creux}\ and\ \citenamefont
  {Harmer}(2019)}]{creux2019circadian}%
  \BibitemOpen
  \bibfield  {author} {\bibinfo {author} {\bibfnamefont {N.}~\bibnamefont
  {Creux}}\ and\ \bibinfo {author} {\bibfnamefont {S.}~\bibnamefont {Harmer}},\
  }\bibfield  {title} {\bibinfo {title} {Circadian rhythms in plants},\ }\href
  {https://doi.org/10.1101/cshperspect.a034611} {\bibfield  {journal} {\bibinfo
   {journal} {Cold Spring Harb. Perspect. Biol.}\ }\textbf {\bibinfo {volume}
  {11}},\ \bibinfo {pages} {a034611} (\bibinfo {year} {2019})}\BibitemShut
  {NoStop}%
\bibitem [{\citenamefont {Engquist}(2016)}]{engquist2016encyclopedia}%
  \BibitemOpen
  \bibfield  {author} {\bibinfo {author} {\bibfnamefont {B.}~\bibnamefont
  {Engquist}},\ }\href {https://doi.org/10.1007/978-3-540-70529-1} {\emph
  {\bibinfo {title} {Encyclopedia of Applied and Computational Mathematics}}}\
  (\bibinfo  {publisher} {Springer Publishing Company, Incorporated},\ \bibinfo
  {year} {2016})\BibitemShut {NoStop}%
\bibitem [{\citenamefont {Baserga}(1968)}]{baserga1968biochemistry}%
  \BibitemOpen
  \bibfield  {author} {\bibinfo {author} {\bibfnamefont {R.}~\bibnamefont
  {Baserga}},\ }\bibfield  {title} {\bibinfo {title} {Biochemistry of the cell
  cycle: a review},\ }\href
  {https://doi.org/10.1111/j.1365-2184.1968.tb00957.x} {\bibfield  {journal}
  {\bibinfo  {journal} {Cell Prolif.}\ }\textbf {\bibinfo {volume} {1}},\
  \bibinfo {pages} {167} (\bibinfo {year} {1968})}\BibitemShut {NoStop}%
\bibitem [{\citenamefont {Schafer}(1998)}]{schafer1998cell}%
  \BibitemOpen
  \bibfield  {author} {\bibinfo {author} {\bibfnamefont {K.}~\bibnamefont
  {Schafer}},\ }\bibfield  {title} {\bibinfo {title} {The cell cycle: a
  review},\ }\href {https://doi.org/10.1177/030098589803500601} {\bibfield
  {journal} {\bibinfo  {journal} {Vet. Pathol.}\ }\textbf {\bibinfo {volume}
  {35}},\ \bibinfo {pages} {461} (\bibinfo {year} {1998})}\BibitemShut
  {NoStop}%
\bibitem [{\citenamefont {Vermeulen}\ \emph {et~al.}(2003)\citenamefont
  {Vermeulen}, \citenamefont {Van~Bockstaele},\ and\ \citenamefont
  {Berneman}}]{vermeulen2003cell}%
  \BibitemOpen
  \bibfield  {author} {\bibinfo {author} {\bibfnamefont {K.}~\bibnamefont
  {Vermeulen}}, \bibinfo {author} {\bibfnamefont {D.~R.}\ \bibnamefont
  {Van~Bockstaele}},\ and\ \bibinfo {author} {\bibfnamefont {Z.~N.}\
  \bibnamefont {Berneman}},\ }\bibfield  {title} {\bibinfo {title} {The cell
  cycle: a review of regulation, deregulation and therapeutic targets in
  cancer},\ }\href {https://doi.org/j.1365-2184.2003.00266.x} {\bibfield
  {journal} {\bibinfo  {journal} {Cell Prolif.}\ }\textbf {\bibinfo {volume}
  {36}},\ \bibinfo {pages} {131} (\bibinfo {year} {2003})}\BibitemShut
  {NoStop}%
\bibitem [{\citenamefont {Golias}\ \emph {et~al.}(2004)\citenamefont {Golias},
  \citenamefont {Charalabopoulos},\ and\ \citenamefont
  {Charalabopoulos}}]{golias2004cell}%
  \BibitemOpen
  \bibfield  {author} {\bibinfo {author} {\bibfnamefont {C.}~\bibnamefont
  {Golias}}, \bibinfo {author} {\bibfnamefont {A.}~\bibnamefont
  {Charalabopoulos}},\ and\ \bibinfo {author} {\bibfnamefont {K.}~\bibnamefont
  {Charalabopoulos}},\ }\bibfield  {title} {\bibinfo {title} {Cell
  proliferation and cell cycle control: a mini review},\ }\href@noop {}
  {\bibfield  {journal} {\bibinfo  {journal} {International journal of clinical
  practice}\ }\textbf {\bibinfo {volume} {58}},\ \bibinfo {pages} {1134}
  (\bibinfo {year} {2004})}\BibitemShut {NoStop}%
\bibitem [{\citenamefont {Alberts}\ \emph {et~al.}(2015)\citenamefont
  {Alberts}, \citenamefont {Bray}, \citenamefont {Hopkin}, \citenamefont
  {Johnson}, \citenamefont {Lewis}, \citenamefont {Raff}, \citenamefont
  {Roberts},\ and\ \citenamefont {Walter}}]{alberts2015essential}%
  \BibitemOpen
  \bibfield  {author} {\bibinfo {author} {\bibfnamefont {B.}~\bibnamefont
  {Alberts}}, \bibinfo {author} {\bibfnamefont {D.}~\bibnamefont {Bray}},
  \bibinfo {author} {\bibfnamefont {K.}~\bibnamefont {Hopkin}}, \bibinfo
  {author} {\bibfnamefont {A.~D.}\ \bibnamefont {Johnson}}, \bibinfo {author}
  {\bibfnamefont {J.}~\bibnamefont {Lewis}}, \bibinfo {author} {\bibfnamefont
  {M.}~\bibnamefont {Raff}}, \bibinfo {author} {\bibfnamefont {K.}~\bibnamefont
  {Roberts}},\ and\ \bibinfo {author} {\bibfnamefont {P.}~\bibnamefont
  {Walter}},\ }\href@noop {} {\emph {\bibinfo {title} {Essential cell
  biology}}}\ (\bibinfo  {publisher} {Garland Science},\ \bibinfo {year}
  {2015})\BibitemShut {NoStop}%
\bibitem [{\citenamefont {Matson}\ and\ \citenamefont
  {Cook}(2017)}]{matson2017cell}%
  \BibitemOpen
  \bibfield  {author} {\bibinfo {author} {\bibfnamefont {J.~P.}\ \bibnamefont
  {Matson}}\ and\ \bibinfo {author} {\bibfnamefont {J.~G.}\ \bibnamefont
  {Cook}},\ }\bibfield  {title} {\bibinfo {title} {Cell cycle proliferation
  decisions: the impact of single cell analyses},\ }\href@noop {} {\bibfield
  {journal} {\bibinfo  {journal} {The FEBS journal}\ }\textbf {\bibinfo
  {volume} {284}},\ \bibinfo {pages} {362} (\bibinfo {year}
  {2017})}\BibitemShut {NoStop}%
\bibitem [{\citenamefont {Li}\ and\ \citenamefont
  {Wang}(2014)}]{li2014landscape}%
  \BibitemOpen
  \bibfield  {author} {\bibinfo {author} {\bibfnamefont {C.}~\bibnamefont
  {Li}}\ and\ \bibinfo {author} {\bibfnamefont {J.}~\bibnamefont {Wang}},\
  }\bibfield  {title} {\bibinfo {title} {Landscape and flux reveal a new global
  view and physical quantification of mammalian cell cycle},\ }\href@noop {}
  {\bibfield  {journal} {\bibinfo  {journal} {Proceedings of the National
  Academy of Sciences}\ }\textbf {\bibinfo {volume} {111}},\ \bibinfo {pages}
  {14130} (\bibinfo {year} {2014})}\BibitemShut {NoStop}%
\bibitem [{\citenamefont {Gao}\ \emph {et~al.}(2017)\citenamefont {Gao},
  \citenamefont {He}, \citenamefont {Shi}, \citenamefont {Cai}, \citenamefont
  {Xu}, \citenamefont {Jiang}, \citenamefont {Zhang},\ and\ \citenamefont
  {Wang}}]{gao2017cell}%
  \BibitemOpen
  \bibfield  {author} {\bibinfo {author} {\bibfnamefont {J.}~\bibnamefont
  {Gao}}, \bibinfo {author} {\bibfnamefont {L.}~\bibnamefont {He}}, \bibinfo
  {author} {\bibfnamefont {Y.}~\bibnamefont {Shi}}, \bibinfo {author}
  {\bibfnamefont {M.}~\bibnamefont {Cai}}, \bibinfo {author} {\bibfnamefont
  {H.}~\bibnamefont {Xu}}, \bibinfo {author} {\bibfnamefont {J.}~\bibnamefont
  {Jiang}}, \bibinfo {author} {\bibfnamefont {L.}~\bibnamefont {Zhang}},\ and\
  \bibinfo {author} {\bibfnamefont {H.}~\bibnamefont {Wang}},\ }\bibfield
  {title} {\bibinfo {title} {Cell contact and pressure control of yap
  localization and clustering revealed by super-resolution imaging},\ }\href
  {https://doi.org/10.1039/c7nr05818g} {\bibfield  {journal} {\bibinfo
  {journal} {Nanoscale}\ }\textbf {\bibinfo {volume} {9}},\ \bibinfo {pages}
  {16993} (\bibinfo {year} {2017})}\BibitemShut {NoStop}%
\bibitem [{\citenamefont {Takao}\ \emph {et~al.}(2019)\citenamefont {Takao},
  \citenamefont {Taya},\ and\ \citenamefont {Chiew}}]{takao2019mechanical}%
  \BibitemOpen
  \bibfield  {author} {\bibinfo {author} {\bibfnamefont {S.}~\bibnamefont
  {Takao}}, \bibinfo {author} {\bibfnamefont {M.}~\bibnamefont {Taya}},\ and\
  \bibinfo {author} {\bibfnamefont {C.}~\bibnamefont {Chiew}},\ }\bibfield
  {title} {\bibinfo {title} {Mechanical stress-induced cell death in breast
  cancer cells},\ }\href {https://doi.org/10.1242/bio.043133} {\bibfield
  {journal} {\bibinfo  {journal} {Biology open}\ }\textbf {\bibinfo {volume}
  {8}},\ \bibinfo {pages} {bio043133} (\bibinfo {year} {2019})}\BibitemShut
  {NoStop}%
\bibitem [{\citenamefont {Wagstaff}\ \emph {et~al.}(2016)\citenamefont
  {Wagstaff}, \citenamefont {Goschorska}, \citenamefont {Kozyrska},
  \citenamefont {Duclos}, \citenamefont {Kucinski}, \citenamefont {Chessel},
  \citenamefont {Hampton-O’Neil}, \citenamefont {Bradshaw}, \citenamefont
  {Allen}, \citenamefont {Rawlins} \emph {et~al.}}]{wagstaff2016mechanical}%
  \BibitemOpen
  \bibfield  {author} {\bibinfo {author} {\bibfnamefont {L.}~\bibnamefont
  {Wagstaff}}, \bibinfo {author} {\bibfnamefont {M.}~\bibnamefont
  {Goschorska}}, \bibinfo {author} {\bibfnamefont {K.}~\bibnamefont
  {Kozyrska}}, \bibinfo {author} {\bibfnamefont {G.}~\bibnamefont {Duclos}},
  \bibinfo {author} {\bibfnamefont {I.}~\bibnamefont {Kucinski}}, \bibinfo
  {author} {\bibfnamefont {A.}~\bibnamefont {Chessel}}, \bibinfo {author}
  {\bibfnamefont {L.}~\bibnamefont {Hampton-O’Neil}}, \bibinfo {author}
  {\bibfnamefont {C.~R.}\ \bibnamefont {Bradshaw}}, \bibinfo {author}
  {\bibfnamefont {G.~E.}\ \bibnamefont {Allen}}, \bibinfo {author}
  {\bibfnamefont {E.~L.}\ \bibnamefont {Rawlins}}, \emph {et~al.},\ }\bibfield
  {title} {\bibinfo {title} {Mechanical cell competition kills cells via
  induction of lethal p53 levels},\ }\href
  {https://doi.org/10.1038/ncomms11373} {\bibfield  {journal} {\bibinfo
  {journal} {Nat. Commun.}\ }\textbf {\bibinfo {volume} {7}},\ \bibinfo {pages}
  {1} (\bibinfo {year} {2016})}\BibitemShut {NoStop}%
\bibitem [{\citenamefont {Mascharak}\ \emph {et~al.}(2017)\citenamefont
  {Mascharak}, \citenamefont {Benitez}, \citenamefont {Proctor}, \citenamefont
  {Madl}, \citenamefont {Hu}, \citenamefont {Dewi}, \citenamefont {Butte},\
  and\ \citenamefont {Heilshorn}}]{mascharak2017yap}%
  \BibitemOpen
  \bibfield  {author} {\bibinfo {author} {\bibfnamefont {S.}~\bibnamefont
  {Mascharak}}, \bibinfo {author} {\bibfnamefont {P.~L.}\ \bibnamefont
  {Benitez}}, \bibinfo {author} {\bibfnamefont {A.~C.}\ \bibnamefont
  {Proctor}}, \bibinfo {author} {\bibfnamefont {C.~M.}\ \bibnamefont {Madl}},
  \bibinfo {author} {\bibfnamefont {K.~H.}\ \bibnamefont {Hu}}, \bibinfo
  {author} {\bibfnamefont {R.~E.}\ \bibnamefont {Dewi}}, \bibinfo {author}
  {\bibfnamefont {M.~J.}\ \bibnamefont {Butte}},\ and\ \bibinfo {author}
  {\bibfnamefont {S.~C.}\ \bibnamefont {Heilshorn}},\ }\bibfield  {title}
  {\bibinfo {title} {Yap-dependent mechanotransduction is required for
  proliferation and migration on native-like substrate topography},\ }\href
  {https://doi.org/10.1016/j.biomaterials.2016.11.019} {\bibfield  {journal}
  {\bibinfo  {journal} {Biomaterials}\ }\textbf {\bibinfo {volume} {115}},\
  \bibinfo {pages} {155} (\bibinfo {year} {2017})}\BibitemShut {NoStop}%
\bibitem [{\citenamefont {Raj}\ and\ \citenamefont
  {Bam}(2019)}]{raj2019reciprocal}%
  \BibitemOpen
  \bibfield  {author} {\bibinfo {author} {\bibfnamefont {N.}~\bibnamefont
  {Raj}}\ and\ \bibinfo {author} {\bibfnamefont {R.}~\bibnamefont {Bam}},\
  }\bibfield  {title} {\bibinfo {title} {Reciprocal crosstalk between
  yap1/hippo pathway and the p53 family proteins: Mechanisms and outcomes in
  cancer},\ }\href {https://doi.org/10.3389/fcell.2019.00159} {\bibfield
  {journal} {\bibinfo  {journal} {Front. Cell Dev. Biol.}\ }\textbf {\bibinfo
  {volume} {7}},\ \bibinfo {pages} {159} (\bibinfo {year} {2019})}\BibitemShut
  {NoStop}%
\bibitem [{\citenamefont {Perez-Gonzalez}\ \emph {et~al.}(2019)\citenamefont
  {Perez-Gonzalez}, \citenamefont {Rochman}, \citenamefont {Yao}, \citenamefont
  {Tao}, \citenamefont {Le}, \citenamefont {Flanary}, \citenamefont {Sablich},
  \citenamefont {Toler}, \citenamefont {Crentsil}, \citenamefont {Takaesu}
  \emph {et~al.}}]{perez2019yap}%
  \BibitemOpen
  \bibfield  {author} {\bibinfo {author} {\bibfnamefont {N.~A.}\ \bibnamefont
  {Perez-Gonzalez}}, \bibinfo {author} {\bibfnamefont {N.~D.}\ \bibnamefont
  {Rochman}}, \bibinfo {author} {\bibfnamefont {K.}~\bibnamefont {Yao}},
  \bibinfo {author} {\bibfnamefont {J.}~\bibnamefont {Tao}}, \bibinfo {author}
  {\bibfnamefont {M.-T.~T.}\ \bibnamefont {Le}}, \bibinfo {author}
  {\bibfnamefont {S.}~\bibnamefont {Flanary}}, \bibinfo {author} {\bibfnamefont
  {L.}~\bibnamefont {Sablich}}, \bibinfo {author} {\bibfnamefont
  {B.}~\bibnamefont {Toler}}, \bibinfo {author} {\bibfnamefont
  {E.}~\bibnamefont {Crentsil}}, \bibinfo {author} {\bibfnamefont
  {F.}~\bibnamefont {Takaesu}}, \emph {et~al.},\ }\bibfield  {title} {\bibinfo
  {title} {Yap and taz regulate cell volume},\ }\href
  {https://doi.org/10.1083/jcb.201902067} {\bibfield  {journal} {\bibinfo
  {journal} {J. Cell Biol.}\ }\textbf {\bibinfo {volume} {218}},\ \bibinfo
  {pages} {3472} (\bibinfo {year} {2019})}\BibitemShut {NoStop}%
\bibitem [{\citenamefont {Vining}\ and\ \citenamefont
  {Mooney}(2017)}]{vining2017mechanical}%
  \BibitemOpen
  \bibfield  {author} {\bibinfo {author} {\bibfnamefont {K.~H.}\ \bibnamefont
  {Vining}}\ and\ \bibinfo {author} {\bibfnamefont {D.~J.}\ \bibnamefont
  {Mooney}},\ }\bibfield  {title} {\bibinfo {title} {Mechanical forces direct
  stem cell behaviour in development and regeneration},\ }\href
  {https://doi.org/10.1038/nrm.2017.108} {\bibfield  {journal} {\bibinfo
  {journal} {Nat. Rev. Microbiol.}\ }\textbf {\bibinfo {volume} {18}},\
  \bibinfo {pages} {728} (\bibinfo {year} {2017})}\BibitemShut {NoStop}%
\bibitem [{\citenamefont {Chu}\ \emph {et~al.}(2018)\citenamefont {Chu},
  \citenamefont {Kilic}, \citenamefont {Cho}, \citenamefont {Groisman},\ and\
  \citenamefont {Levchenko}}]{chu2018self}%
  \BibitemOpen
  \bibfield  {author} {\bibinfo {author} {\bibfnamefont {E.~K.}\ \bibnamefont
  {Chu}}, \bibinfo {author} {\bibfnamefont {O.}~\bibnamefont {Kilic}}, \bibinfo
  {author} {\bibfnamefont {H.}~\bibnamefont {Cho}}, \bibinfo {author}
  {\bibfnamefont {A.}~\bibnamefont {Groisman}},\ and\ \bibinfo {author}
  {\bibfnamefont {A.}~\bibnamefont {Levchenko}},\ }\bibfield  {title} {\bibinfo
  {title} {Self-induced mechanical stress can trigger biofilm formation in
  uropathogenic escherichia coli},\ }\href
  {https://doi.org/10.1038/s41467-018-06552-z} {\bibfield  {journal} {\bibinfo
  {journal} {Nat. Commun.}\ }\textbf {\bibinfo {volume} {9}},\ \bibinfo {pages}
  {1} (\bibinfo {year} {2018})}\BibitemShut {NoStop}%
\bibitem [{\citenamefont {Saeed}\ and\ \citenamefont
  {Weihs}(2019)}]{saeed2019finite}%
  \BibitemOpen
  \bibfield  {author} {\bibinfo {author} {\bibfnamefont {M.}~\bibnamefont
  {Saeed}}\ and\ \bibinfo {author} {\bibfnamefont {D.}~\bibnamefont {Weihs}},\
  }\bibfield  {title} {\bibinfo {title} {Finite element analysis reveals an
  important role for cell morphology in response to mechanical compression},\
  }\href {https://doi.org/10.1007/s10237-019-01276-5} {\bibfield  {journal}
  {\bibinfo  {journal} {Biomech. Model. Mechanobiol.}\ ,\ \bibinfo {pages} {1}}
  (\bibinfo {year} {2019})}\BibitemShut {NoStop}%
\bibitem [{\citenamefont {Levayer}(2020)}]{levayer2020solid}%
  \BibitemOpen
  \bibfield  {author} {\bibinfo {author} {\bibfnamefont {R.}~\bibnamefont
  {Levayer}},\ }\bibfield  {title} {\bibinfo {title} {Solid stress, competition
  for space and cancer: The opposing roles of mechanical cell competition in
  tumour initiation and growth},\ }\href
  {https://doi.org/10.1016/j.semcancer.2019.05.004} {\bibfield  {journal}
  {\bibinfo  {journal} {Semin. Cancer Biol.}\ }\textbf {\bibinfo {volume}
  {63}},\ \bibinfo {pages} {69} (\bibinfo {year} {2020})}\BibitemShut {NoStop}%
\bibitem [{\citenamefont {Alt}\ \emph {et~al.}(2017)\citenamefont {Alt},
  \citenamefont {Ganguly},\ and\ \citenamefont {Salbreux}}]{alt2017vertex}%
  \BibitemOpen
  \bibfield  {author} {\bibinfo {author} {\bibfnamefont {S.}~\bibnamefont
  {Alt}}, \bibinfo {author} {\bibfnamefont {P.}~\bibnamefont {Ganguly}},\ and\
  \bibinfo {author} {\bibfnamefont {G.}~\bibnamefont {Salbreux}},\ }\bibfield
  {title} {\bibinfo {title} {Vertex models: from cell mechanics to tissue
  morphogenesis},\ }\href {https://doi.org/10.1098/rstb.2015.0520} {\bibfield
  {journal} {\bibinfo  {journal} {Phil. Trans. R. Soc. B}\ }\textbf {\bibinfo
  {volume} {372}},\ \bibinfo {pages} {20150520} (\bibinfo {year}
  {2017})}\BibitemShut {NoStop}%
\bibitem [{\citenamefont {Alert}\ and\ \citenamefont
  {Trepat}(2020)}]{alert2020physical}%
  \BibitemOpen
  \bibfield  {author} {\bibinfo {author} {\bibfnamefont {R.}~\bibnamefont
  {Alert}}\ and\ \bibinfo {author} {\bibfnamefont {X.}~\bibnamefont {Trepat}},\
  }\bibfield  {title} {\bibinfo {title} {Physical models of collective cell
  migration},\ }\href
  {https://doi.org/10.1146/annurev-conmatphys-031218-013516} {\bibfield
  {journal} {\bibinfo  {journal} {Annu. Rev. Condens. Matter Phys.}\ }\textbf
  {\bibinfo {volume} {11}},\ \bibinfo {pages} {77} (\bibinfo {year}
  {2020})}\BibitemShut {NoStop}%
\bibitem [{\citenamefont {Blanch-Mercader}\ \emph {et~al.}(2017)\citenamefont
  {Blanch-Mercader}, \citenamefont {Vincent}, \citenamefont
  {Bazelli{\`{e}}res}, \citenamefont {Serra-Picamal}, \citenamefont {Trepat},\
  and\ \citenamefont {Casademunt}}]{Blanch-Mercader2017}%
  \BibitemOpen
  \bibfield  {author} {\bibinfo {author} {\bibfnamefont {C.}~\bibnamefont
  {Blanch-Mercader}}, \bibinfo {author} {\bibfnamefont {R.}~\bibnamefont
  {Vincent}}, \bibinfo {author} {\bibfnamefont {E.}~\bibnamefont
  {Bazelli{\`{e}}res}}, \bibinfo {author} {\bibfnamefont {X.}~\bibnamefont
  {Serra-Picamal}}, \bibinfo {author} {\bibfnamefont {X.}~\bibnamefont
  {Trepat}},\ and\ \bibinfo {author} {\bibfnamefont {J.}~\bibnamefont
  {Casademunt}},\ }\bibfield  {title} {\bibinfo {title} {{Effective viscosity
  and dynamics of spreading epithelia: a solvable model}},\ }\href
  {https://doi.org/10.1039/C6SM02188C} {\bibfield  {journal} {\bibinfo
  {journal} {Soft Matter}\ }\textbf {\bibinfo {volume} {13}},\ \bibinfo {pages}
  {1235} (\bibinfo {year} {2017})}\BibitemShut {NoStop}%
\bibitem [{\citenamefont {Ishihara}\ \emph {et~al.}(2017)\citenamefont
  {Ishihara}, \citenamefont {Marcq},\ and\ \citenamefont
  {Sugimura}}]{Ishihara2017}%
  \BibitemOpen
  \bibfield  {author} {\bibinfo {author} {\bibfnamefont {S.}~\bibnamefont
  {Ishihara}}, \bibinfo {author} {\bibfnamefont {P.}~\bibnamefont {Marcq}},\
  and\ \bibinfo {author} {\bibfnamefont {K.}~\bibnamefont {Sugimura}},\
  }\bibfield  {title} {\bibinfo {title} {From cells to tissue: A continuum
  model of epithelial mechanics},\ }\href
  {https://doi.org/10.1103/PhysRevE.96.022418} {\bibfield  {journal} {\bibinfo
  {journal} {Physical Review E}\ }\textbf {\bibinfo {volume} {96}},\ \bibinfo
  {pages} {022418} (\bibinfo {year} {2017})}\BibitemShut {NoStop}%
\bibitem [{\citenamefont {J{\"u}licher}\ \emph {et~al.}(2018)\citenamefont
  {J{\"u}licher}, \citenamefont {Grill},\ and\ \citenamefont
  {Salbreux}}]{julicher2018hydrodynamic}%
  \BibitemOpen
  \bibfield  {author} {\bibinfo {author} {\bibfnamefont {F.}~\bibnamefont
  {J{\"u}licher}}, \bibinfo {author} {\bibfnamefont {S.~W.}\ \bibnamefont
  {Grill}},\ and\ \bibinfo {author} {\bibfnamefont {G.}~\bibnamefont
  {Salbreux}},\ }\bibfield  {title} {\bibinfo {title} {Hydrodynamic theory of
  active matter},\ }\href {https://doi.org/10.1088/1361-6633/aab6bb} {\bibfield
   {journal} {\bibinfo  {journal} {Rep. Prog. Phys.}\ }\textbf {\bibinfo
  {volume} {81}},\ \bibinfo {pages} {076601} (\bibinfo {year}
  {2018})}\BibitemShut {NoStop}%
\bibitem [{\citenamefont {Williamson}\ and\ \citenamefont
  {Salbreux}(2018)}]{williamson2018stability}%
  \BibitemOpen
  \bibfield  {author} {\bibinfo {author} {\bibfnamefont {J.~J.}\ \bibnamefont
  {Williamson}}\ and\ \bibinfo {author} {\bibfnamefont {G.}~\bibnamefont
  {Salbreux}},\ }\bibfield  {title} {\bibinfo {title} {Stability and roughness
  of interfaces in mechanically regulated tissues},\ }\href
  {https://doi.org/10.1103/PhysRevLett.121.238102} {\bibfield  {journal}
  {\bibinfo  {journal} {Phys. Rev. Lett.}\ }\textbf {\bibinfo {volume} {121}},\
  \bibinfo {pages} {238102} (\bibinfo {year} {2018})}\BibitemShut {NoStop}%
\bibitem [{\citenamefont {Banerjee}\ and\ \citenamefont
  {Marchetti}(2019)}]{banerjee2019continuum}%
  \BibitemOpen
  \bibfield  {author} {\bibinfo {author} {\bibfnamefont {S.}~\bibnamefont
  {Banerjee}}\ and\ \bibinfo {author} {\bibfnamefont {M.~C.}\ \bibnamefont
  {Marchetti}},\ }\bibfield  {title} {\bibinfo {title} {Continuum models of
  collective cell migration},\ }in\ \href
  {https://doi.org/10.1007/978-3-030-17593-1_4} {\emph {\bibinfo {booktitle}
  {Cell Migrations: Causes and Functions}}}\ (\bibinfo  {publisher}
  {Springer},\ \bibinfo {year} {2019})\ pp.\ \bibinfo {pages}
  {45--66}\BibitemShut {NoStop}%
\bibitem [{\citenamefont {Xi}\ \emph {et~al.}(2019)\citenamefont {Xi},
  \citenamefont {Saw}, \citenamefont {Delacour}, \citenamefont {Lim},\ and\
  \citenamefont {Ladoux}}]{xi2019material}%
  \BibitemOpen
  \bibfield  {author} {\bibinfo {author} {\bibfnamefont {W.}~\bibnamefont
  {Xi}}, \bibinfo {author} {\bibfnamefont {T.~B.}\ \bibnamefont {Saw}},
  \bibinfo {author} {\bibfnamefont {D.}~\bibnamefont {Delacour}}, \bibinfo
  {author} {\bibfnamefont {C.~T.}\ \bibnamefont {Lim}},\ and\ \bibinfo {author}
  {\bibfnamefont {B.}~\bibnamefont {Ladoux}},\ }\bibfield  {title} {\bibinfo
  {title} {Material approaches to active tissue mechanics},\ }\href
  {https://doi.org/10.1038/s41578-018-0066-z} {\bibfield  {journal} {\bibinfo
  {journal} {Nat. Rev. Mater.}\ }\textbf {\bibinfo {volume} {4}},\ \bibinfo
  {pages} {23} (\bibinfo {year} {2019})}\BibitemShut {NoStop}%
\bibitem [{\citenamefont {Schnyder}\ \emph {et~al.}(2020)\citenamefont
  {Schnyder}, \citenamefont {Molina},\ and\ \citenamefont
  {Yamamoto}}]{schnyder2020control}%
  \BibitemOpen
  \bibfield  {author} {\bibinfo {author} {\bibfnamefont {S.~K.}\ \bibnamefont
  {Schnyder}}, \bibinfo {author} {\bibfnamefont {J.~J.}\ \bibnamefont
  {Molina}},\ and\ \bibinfo {author} {\bibfnamefont {R.}~\bibnamefont
  {Yamamoto}},\ }\bibfield  {title} {\bibinfo {title} {Control of cell colony
  growth by contact inhibition},\ }\href
  {https://doi.org/10.1038/s41598-020-62913-z} {\bibfield  {journal} {\bibinfo
  {journal} {Sci. Rep.}\ }\textbf {\bibinfo {volume} {10}},\ \bibinfo {pages}
  {1} (\bibinfo {year} {2020})}\BibitemShut {NoStop}%
\bibitem [{\citenamefont {Novak}\ \emph {et~al.}(2007)\citenamefont {Novak},
  \citenamefont {Tyson}, \citenamefont {Gyorffy},\ and\ \citenamefont
  {Csikasz-Nagy}}]{novak2007irreversible}%
  \BibitemOpen
  \bibfield  {author} {\bibinfo {author} {\bibfnamefont {B.}~\bibnamefont
  {Novak}}, \bibinfo {author} {\bibfnamefont {J.~J.}\ \bibnamefont {Tyson}},
  \bibinfo {author} {\bibfnamefont {B.}~\bibnamefont {Gyorffy}},\ and\ \bibinfo
  {author} {\bibfnamefont {A.}~\bibnamefont {Csikasz-Nagy}},\ }\bibfield
  {title} {\bibinfo {title} {Irreversible cell-cycle transitions are due to
  systems-level feedback},\ }\href {https://doi.org/10.1038/ncb0707-724}
  {\bibfield  {journal} {\bibinfo  {journal} {Nature cell biology}\ }\textbf
  {\bibinfo {volume} {9}},\ \bibinfo {pages} {724} (\bibinfo {year}
  {2007})}\BibitemShut {NoStop}%
\bibitem [{\citenamefont {L{\'o}pez-Avil{\'e}s}\ \emph
  {et~al.}(2009)\citenamefont {L{\'o}pez-Avil{\'e}s}, \citenamefont {Kapuy},
  \citenamefont {Nov{\'a}k},\ and\ \citenamefont
  {Uhlmann}}]{lopez2009irreversibility}%
  \BibitemOpen
  \bibfield  {author} {\bibinfo {author} {\bibfnamefont {S.}~\bibnamefont
  {L{\'o}pez-Avil{\'e}s}}, \bibinfo {author} {\bibfnamefont {O.}~\bibnamefont
  {Kapuy}}, \bibinfo {author} {\bibfnamefont {B.}~\bibnamefont {Nov{\'a}k}},\
  and\ \bibinfo {author} {\bibfnamefont {F.}~\bibnamefont {Uhlmann}},\
  }\bibfield  {title} {\bibinfo {title} {Irreversibility of mitotic exit is the
  consequence of systems-level feedback},\ }\href
  {https://doi.org/10.1038/nature07984} {\bibfield  {journal} {\bibinfo
  {journal} {Nature}\ }\textbf {\bibinfo {volume} {459}},\ \bibinfo {pages}
  {592} (\bibinfo {year} {2009})}\BibitemShut {NoStop}%
\bibitem [{\citenamefont {Streichan}\ \emph {et~al.}(2014)\citenamefont
  {Streichan}, \citenamefont {Hoerner}, \citenamefont {Schneidt}, \citenamefont
  {Holzer},\ and\ \citenamefont {Hufnagel}}]{streichan2014spatial}%
  \BibitemOpen
  \bibfield  {author} {\bibinfo {author} {\bibfnamefont {S.~J.}\ \bibnamefont
  {Streichan}}, \bibinfo {author} {\bibfnamefont {C.~R.}\ \bibnamefont
  {Hoerner}}, \bibinfo {author} {\bibfnamefont {T.}~\bibnamefont {Schneidt}},
  \bibinfo {author} {\bibfnamefont {D.}~\bibnamefont {Holzer}},\ and\ \bibinfo
  {author} {\bibfnamefont {L.}~\bibnamefont {Hufnagel}},\ }\bibfield  {title}
  {\bibinfo {title} {Spatial constraints control cell proliferation in
  tissues},\ }\href {https://doi.org/10.1073/pnas.1323016111} {\bibfield
  {journal} {\bibinfo  {journal} {Proceedings of the National Academy of
  Sciences}\ }\textbf {\bibinfo {volume} {111}},\ \bibinfo {pages} {5586}
  (\bibinfo {year} {2014})}\BibitemShut {NoStop}%
\bibitem [{\citenamefont {Duszyc}\ \emph {et~al.}(2018)\citenamefont {Duszyc},
  \citenamefont {Viasnoff} \emph {et~al.}}]{duszyc2018mechanosensing}%
  \BibitemOpen
  \bibfield  {author} {\bibinfo {author} {\bibfnamefont {K.}~\bibnamefont
  {Duszyc}}, \bibinfo {author} {\bibfnamefont {V.}~\bibnamefont {Viasnoff}},
  \emph {et~al.},\ }\bibfield  {title} {\bibinfo {title} {Mechanosensing and
  mechanotransduction at cell--cell junctions},\ }\href
  {https://doi.org/10.1101/cshperspect.a028761} {\bibfield  {journal} {\bibinfo
   {journal} {Cold Spring Harbor perspectives in biology}\ }\textbf {\bibinfo
  {volume} {10}},\ \bibinfo {pages} {a028761} (\bibinfo {year}
  {2018})}\BibitemShut {NoStop}%
\bibitem [{\citenamefont {Rizzuti}\ \emph {et~al.}(2020)\citenamefont
  {Rizzuti}, \citenamefont {Mascheroni}, \citenamefont {Arcucci}, \citenamefont
  {Ben-M{\'e}riem}, \citenamefont {Prunet}, \citenamefont {Barentin},
  \citenamefont {Rivi{\`e}re}, \citenamefont {Delano{\"e}-Ayari}, \citenamefont
  {Hatzikirou}, \citenamefont {Guillermet-Guibert} \emph
  {et~al.}}]{rizzuti2020mechanical}%
  \BibitemOpen
  \bibfield  {author} {\bibinfo {author} {\bibfnamefont {I.~F.}\ \bibnamefont
  {Rizzuti}}, \bibinfo {author} {\bibfnamefont {P.}~\bibnamefont {Mascheroni}},
  \bibinfo {author} {\bibfnamefont {S.}~\bibnamefont {Arcucci}}, \bibinfo
  {author} {\bibfnamefont {Z.}~\bibnamefont {Ben-M{\'e}riem}}, \bibinfo
  {author} {\bibfnamefont {A.}~\bibnamefont {Prunet}}, \bibinfo {author}
  {\bibfnamefont {C.}~\bibnamefont {Barentin}}, \bibinfo {author}
  {\bibfnamefont {C.}~\bibnamefont {Rivi{\`e}re}}, \bibinfo {author}
  {\bibfnamefont {H.}~\bibnamefont {Delano{\"e}-Ayari}}, \bibinfo {author}
  {\bibfnamefont {H.}~\bibnamefont {Hatzikirou}}, \bibinfo {author}
  {\bibfnamefont {J.}~\bibnamefont {Guillermet-Guibert}}, \emph {et~al.},\
  }\bibfield  {title} {\bibinfo {title} {Mechanical control of cell
  proliferation increases resistance to chemotherapeutic agents},\ }\href
  {https://doi.org/10.1103/PhysRevLett.125.128103} {\bibfield  {journal}
  {\bibinfo  {journal} {Physical Review Letters}\ }\textbf {\bibinfo {volume}
  {125}},\ \bibinfo {pages} {128103} (\bibinfo {year} {2020})}\BibitemShut
  {NoStop}%
\bibitem [{\citenamefont {Ginzberg}\ \emph {et~al.}(2018)\citenamefont
  {Ginzberg}, \citenamefont {Chang}, \citenamefont {D'Souza}, \citenamefont
  {Patel}, \citenamefont {Kafri},\ and\ \citenamefont
  {Kirschner}}]{ginzberg2018cell}%
  \BibitemOpen
  \bibfield  {author} {\bibinfo {author} {\bibfnamefont {M.~B.}\ \bibnamefont
  {Ginzberg}}, \bibinfo {author} {\bibfnamefont {N.}~\bibnamefont {Chang}},
  \bibinfo {author} {\bibfnamefont {H.}~\bibnamefont {D'Souza}}, \bibinfo
  {author} {\bibfnamefont {N.}~\bibnamefont {Patel}}, \bibinfo {author}
  {\bibfnamefont {R.}~\bibnamefont {Kafri}},\ and\ \bibinfo {author}
  {\bibfnamefont {M.~W.}\ \bibnamefont {Kirschner}},\ }\bibfield  {title}
  {\bibinfo {title} {Cell size sensing in animal cells coordinates anabolic
  growth rates and cell cycle progression to maintain cell size uniformity},\
  }\href@noop {} {\bibfield  {journal} {\bibinfo  {journal} {Elife}\ }\textbf
  {\bibinfo {volume} {7}},\ \bibinfo {pages} {e26957} (\bibinfo {year}
  {2018})}\BibitemShut {NoStop}%
\bibitem [{\citenamefont {Vargas-Garcia}\ \emph {et~al.}(2018)\citenamefont
  {Vargas-Garcia}, \citenamefont {Ghusinga},\ and\ \citenamefont
  {Singh}}]{vargas2018cell}%
  \BibitemOpen
  \bibfield  {author} {\bibinfo {author} {\bibfnamefont {C.~A.}\ \bibnamefont
  {Vargas-Garcia}}, \bibinfo {author} {\bibfnamefont {K.~R.}\ \bibnamefont
  {Ghusinga}},\ and\ \bibinfo {author} {\bibfnamefont {A.}~\bibnamefont
  {Singh}},\ }\bibfield  {title} {\bibinfo {title} {Cell size control and gene
  expression homeostasis in single-cells},\ }\href@noop {} {\bibfield
  {journal} {\bibinfo  {journal} {Current opinion in systems biology}\ }\textbf
  {\bibinfo {volume} {8}},\ \bibinfo {pages} {109} (\bibinfo {year}
  {2018})}\BibitemShut {NoStop}%
\bibitem [{\citenamefont {Engeland}(2018)}]{engeland2018cell}%
  \BibitemOpen
  \bibfield  {author} {\bibinfo {author} {\bibfnamefont {K.}~\bibnamefont
  {Engeland}},\ }\bibfield  {title} {\bibinfo {title} {Cell cycle arrest
  through indirect transcriptional repression by p53: I have a dream},\
  }\href@noop {} {\bibfield  {journal} {\bibinfo  {journal} {Cell Death \&
  Differentiation}\ }\textbf {\bibinfo {volume} {25}},\ \bibinfo {pages} {114}
  (\bibinfo {year} {2018})}\BibitemShut {NoStop}%
\bibitem [{\citenamefont {Perez~Gonzalez}\ \emph {et~al.}(2018)\citenamefont
  {Perez~Gonzalez}, \citenamefont {Tao}, \citenamefont {Rochman}, \citenamefont
  {Vig}, \citenamefont {Chiu}, \citenamefont {Wirtz},\ and\ \citenamefont
  {Sun}}]{perez2018cell}%
  \BibitemOpen
  \bibfield  {author} {\bibinfo {author} {\bibfnamefont {N.}~\bibnamefont
  {Perez~Gonzalez}}, \bibinfo {author} {\bibfnamefont {J.}~\bibnamefont {Tao}},
  \bibinfo {author} {\bibfnamefont {N.~D.}\ \bibnamefont {Rochman}}, \bibinfo
  {author} {\bibfnamefont {D.}~\bibnamefont {Vig}}, \bibinfo {author}
  {\bibfnamefont {E.}~\bibnamefont {Chiu}}, \bibinfo {author} {\bibfnamefont
  {D.}~\bibnamefont {Wirtz}},\ and\ \bibinfo {author} {\bibfnamefont {S.~X.}\
  \bibnamefont {Sun}},\ }\bibfield  {title} {\bibinfo {title} {Cell tension and
  mechanical regulation of cell volume},\ }\href@noop {} {\bibfield  {journal}
  {\bibinfo  {journal} {Molecular biology of the cell}\ }\textbf {\bibinfo
  {volume} {29}} (\bibinfo {year} {2018})}\BibitemShut {NoStop}%
\bibitem [{\citenamefont {Zausch}\ and\ \citenamefont
  {Horbach}(2009)}]{zausch2009build}%
  \BibitemOpen
  \bibfield  {author} {\bibinfo {author} {\bibfnamefont {J.}~\bibnamefont
  {Zausch}}\ and\ \bibinfo {author} {\bibfnamefont {J.}~\bibnamefont
  {Horbach}},\ }\bibfield  {title} {\bibinfo {title} {The build-up and
  relaxation of stresses in a glass-forming soft-sphere mixture under shear: A
  computer simulation study},\ }\href
  {https://doi.org/10.1209/0295-5075/88/60001} {\bibfield  {journal} {\bibinfo
  {journal} {EPL}\ }\textbf {\bibinfo {volume} {88}},\ \bibinfo {pages} {60001}
  (\bibinfo {year} {2009})}\BibitemShut {NoStop}%
\bibitem [{\citenamefont {Landau}\ and\ \citenamefont
  {Lifshitz}(1987)}]{Landau1987Fluid}%
  \BibitemOpen
  \bibfield  {author} {\bibinfo {author} {\bibfnamefont {L.~D.}\ \bibnamefont
  {Landau}}\ and\ \bibinfo {author} {\bibfnamefont {E.~M.}\ \bibnamefont
  {Lifshitz}},\ }\href {http://www.worldcat.org/isbn/0750627670} {\emph
  {\bibinfo {title} {Fluid Mechanics}}},\ \bibinfo {edition} {2nd}\ ed.\
  (\bibinfo  {publisher} {Butterworth-Heinemann},\ \bibinfo {year}
  {1987})\BibitemShut {NoStop}%
\bibitem [{\citenamefont {Boyer}\ \emph {et~al.}(2011)\citenamefont {Boyer},
  \citenamefont {Mather}, \citenamefont {Mondrag{\'o}n-Palomino}, \citenamefont
  {Orozco-Fuentes}, \citenamefont {Danino}, \citenamefont {Hasty},\ and\
  \citenamefont {Tsimring}}]{boyer2011buckling}%
  \BibitemOpen
  \bibfield  {author} {\bibinfo {author} {\bibfnamefont {D.}~\bibnamefont
  {Boyer}}, \bibinfo {author} {\bibfnamefont {W.}~\bibnamefont {Mather}},
  \bibinfo {author} {\bibfnamefont {O.}~\bibnamefont {Mondrag{\'o}n-Palomino}},
  \bibinfo {author} {\bibfnamefont {S.}~\bibnamefont {Orozco-Fuentes}},
  \bibinfo {author} {\bibfnamefont {T.}~\bibnamefont {Danino}}, \bibinfo
  {author} {\bibfnamefont {J.}~\bibnamefont {Hasty}},\ and\ \bibinfo {author}
  {\bibfnamefont {L.~S.}\ \bibnamefont {Tsimring}},\ }\bibfield  {title}
  {\bibinfo {title} {Buckling instability in ordered bacterial colonies},\
  }\href {https://doi.org/10.1088/1478-3975/8/2/026008} {\bibfield  {journal}
  {\bibinfo  {journal} {Phys. Biol.}\ }\textbf {\bibinfo {volume} {8}},\
  \bibinfo {pages} {026008} (\bibinfo {year} {2011})}\BibitemShut {NoStop}%
\bibitem [{\citenamefont {Delarue}\ \emph {et~al.}(2013)\citenamefont
  {Delarue}, \citenamefont {Montel}, \citenamefont {Caen}, \citenamefont
  {Elgeti}, \citenamefont {Siaugue}, \citenamefont {Vignjevic}, \citenamefont
  {Prost}, \citenamefont {Joanny},\ and\ \citenamefont
  {Cappello}}]{delarue2013mechanical}%
  \BibitemOpen
  \bibfield  {author} {\bibinfo {author} {\bibfnamefont {M.}~\bibnamefont
  {Delarue}}, \bibinfo {author} {\bibfnamefont {F.}~\bibnamefont {Montel}},
  \bibinfo {author} {\bibfnamefont {O.}~\bibnamefont {Caen}}, \bibinfo {author}
  {\bibfnamefont {J.}~\bibnamefont {Elgeti}}, \bibinfo {author} {\bibfnamefont
  {J.-M.}\ \bibnamefont {Siaugue}}, \bibinfo {author} {\bibfnamefont
  {D.}~\bibnamefont {Vignjevic}}, \bibinfo {author} {\bibfnamefont
  {J.}~\bibnamefont {Prost}}, \bibinfo {author} {\bibfnamefont {J.-F.}\
  \bibnamefont {Joanny}},\ and\ \bibinfo {author} {\bibfnamefont
  {G.}~\bibnamefont {Cappello}},\ }\bibfield  {title} {\bibinfo {title}
  {Mechanical control of cell flow in multicellular spheroids},\ }\href
  {https://doi.org/10.1103/PhysRevLett.110.138103} {\bibfield  {journal}
  {\bibinfo  {journal} {Phys. Rev. Lett.}\ }\textbf {\bibinfo {volume} {110}},\
  \bibinfo {pages} {138103} (\bibinfo {year} {2013})}\BibitemShut {NoStop}%
\bibitem [{\citenamefont {Dolega}\ \emph {et~al.}(2017)\citenamefont {Dolega},
  \citenamefont {Delarue}, \citenamefont {Ingremeau}, \citenamefont {Prost},
  \citenamefont {Delon},\ and\ \citenamefont {Cappello}}]{dolega2017cell}%
  \BibitemOpen
  \bibfield  {author} {\bibinfo {author} {\bibfnamefont {M.}~\bibnamefont
  {Dolega}}, \bibinfo {author} {\bibfnamefont {M.}~\bibnamefont {Delarue}},
  \bibinfo {author} {\bibfnamefont {F.}~\bibnamefont {Ingremeau}}, \bibinfo
  {author} {\bibfnamefont {J.}~\bibnamefont {Prost}}, \bibinfo {author}
  {\bibfnamefont {A.}~\bibnamefont {Delon}},\ and\ \bibinfo {author}
  {\bibfnamefont {G.}~\bibnamefont {Cappello}},\ }\bibfield  {title} {\bibinfo
  {title} {Cell-like pressure sensors reveal increase of mechanical stress
  towards the core of multicellular spheroids under compression},\ }\href
  {https://doi.org/10.1038/ncomms14056} {\bibfield  {journal} {\bibinfo
  {journal} {Nature communications}\ }\textbf {\bibinfo {volume} {8}},\
  \bibinfo {pages} {1} (\bibinfo {year} {2017})}\BibitemShut {NoStop}%
\bibitem [{\citenamefont {Petridou}\ \emph {et~al.}(2017)\citenamefont
  {Petridou}, \citenamefont {Spir{\'o}},\ and\ \citenamefont
  {Heisenberg}}]{petridou2017}%
  \BibitemOpen
  \bibfield  {author} {\bibinfo {author} {\bibfnamefont {N.~I.}\ \bibnamefont
  {Petridou}}, \bibinfo {author} {\bibfnamefont {Z.}~\bibnamefont
  {Spir{\'o}}},\ and\ \bibinfo {author} {\bibfnamefont {C.-P.}\ \bibnamefont
  {Heisenberg}},\ }\bibfield  {title} {\bibinfo {title} {Multiscale force
  sensing in development},\ }\href {https://doi.org/10.1038/ncb3524} {\bibfield
   {journal} {\bibinfo  {journal} {Nature Cell Biology}\ }\textbf {\bibinfo
  {volume} {19}},\ \bibinfo {pages} {581} (\bibinfo {year} {2017})}\BibitemShut
  {NoStop}%
\bibitem [{\citenamefont {Uroz}\ \emph
  {et~al.}(2018{\natexlab{b}})\citenamefont {Uroz}, \citenamefont {Wistorf},
  \citenamefont {Serra-Picamal}, \citenamefont {Conte}, \citenamefont
  {Sales-Pardo}, \citenamefont {Roca-Cusachs}, \citenamefont {Guimer{\`{a}}},\
  and\ \citenamefont {Trepat}}]{Uroz2018}%
  \BibitemOpen
  \bibfield  {author} {\bibinfo {author} {\bibfnamefont {M.}~\bibnamefont
  {Uroz}}, \bibinfo {author} {\bibfnamefont {S.}~\bibnamefont {Wistorf}},
  \bibinfo {author} {\bibfnamefont {X.}~\bibnamefont {Serra-Picamal}}, \bibinfo
  {author} {\bibfnamefont {V.}~\bibnamefont {Conte}}, \bibinfo {author}
  {\bibfnamefont {M.}~\bibnamefont {Sales-Pardo}}, \bibinfo {author}
  {\bibfnamefont {P.}~\bibnamefont {Roca-Cusachs}}, \bibinfo {author}
  {\bibfnamefont {R.}~\bibnamefont {Guimer{\`{a}}}},\ and\ \bibinfo {author}
  {\bibfnamefont {X.}~\bibnamefont {Trepat}},\ }\bibfield  {title} {\bibinfo
  {title} {{Regulation of cell cycle progression by cell-cell and cell-matrix
  forces}},\ }\href {https://doi.org/10.1038/s41556-018-0107-2} {\bibfield
  {journal} {\bibinfo  {journal} {Nature Cell Biology}\ }\textbf {\bibinfo
  {volume} {20}},\ \bibinfo {pages} {646} (\bibinfo {year}
  {2018}{\natexlab{b}})}\BibitemShut {NoStop}%
\bibitem [{\citenamefont {Radmaneshfar}\ \emph {et~al.}(2013)\citenamefont
  {Radmaneshfar}, \citenamefont {Kaloriti}, \citenamefont {Gustin},
  \citenamefont {Gow}, \citenamefont {Brown}, \citenamefont {Grebogi},
  \citenamefont {Romano},\ and\ \citenamefont {Thiel}}]{Radmaneshfar2013}%
  \BibitemOpen
  \bibfield  {author} {\bibinfo {author} {\bibfnamefont {E.}~\bibnamefont
  {Radmaneshfar}}, \bibinfo {author} {\bibfnamefont {D.}~\bibnamefont
  {Kaloriti}}, \bibinfo {author} {\bibfnamefont {M.~C.}\ \bibnamefont
  {Gustin}}, \bibinfo {author} {\bibfnamefont {N.~A.}\ \bibnamefont {Gow}},
  \bibinfo {author} {\bibfnamefont {A.~J.}\ \bibnamefont {Brown}}, \bibinfo
  {author} {\bibfnamefont {C.}~\bibnamefont {Grebogi}}, \bibinfo {author}
  {\bibfnamefont {M.~C.}\ \bibnamefont {Romano}},\ and\ \bibinfo {author}
  {\bibfnamefont {M.}~\bibnamefont {Thiel}},\ }\bibfield  {title} {\bibinfo
  {title} {{From START to FINISH: The Influence of Osmotic Stress on the Cell
  Cycle}},\ }\href {https://doi.org/10.1371/journal.pone.0068067} {\bibfield
  {journal} {\bibinfo  {journal} {PLoS ONE}\ }\textbf {\bibinfo {volume} {8}},\
  \bibinfo {pages} {1} (\bibinfo {year} {2013})}\BibitemShut {NoStop}%
\bibitem [{\citenamefont {Nam}\ \emph {et~al.}(2019)\citenamefont {Nam},
  \citenamefont {Gupta}, \citenamefont {pyo Lee}, \citenamefont {Lee},
  \citenamefont {Wisdom}, \citenamefont {Varma}, \citenamefont {Flaum},
  \citenamefont {Davis}, \citenamefont {West},\ and\ \citenamefont
  {Chaudhuri}}]{Nam2019}%
  \BibitemOpen
  \bibfield  {author} {\bibinfo {author} {\bibfnamefont {S.}~\bibnamefont
  {Nam}}, \bibinfo {author} {\bibfnamefont {V.~K.}\ \bibnamefont {Gupta}},
  \bibinfo {author} {\bibfnamefont {H.}~\bibnamefont {pyo Lee}}, \bibinfo
  {author} {\bibfnamefont {J.~Y.}\ \bibnamefont {Lee}}, \bibinfo {author}
  {\bibfnamefont {K.~M.}\ \bibnamefont {Wisdom}}, \bibinfo {author}
  {\bibfnamefont {S.}~\bibnamefont {Varma}}, \bibinfo {author} {\bibfnamefont
  {E.~M.}\ \bibnamefont {Flaum}}, \bibinfo {author} {\bibfnamefont
  {C.}~\bibnamefont {Davis}}, \bibinfo {author} {\bibfnamefont {R.~B.}\
  \bibnamefont {West}},\ and\ \bibinfo {author} {\bibfnamefont
  {O.}~\bibnamefont {Chaudhuri}},\ }\bibfield  {title} {\bibinfo {title} {{Cell
  cycle progression in confining microenvironments is regulated by a
  growth-responsive TRPV4-PI3K/Akt-p27Kip1 signaling axis}},\ }\bibfield
  {journal} {\bibinfo  {journal} {Science Advances}\ }\textbf {\bibinfo
  {volume} {5}},\ \href {https://doi.org/10.1126/sciadv.aaw6171}
  {10.1126/sciadv.aaw6171} (\bibinfo {year} {2019})\BibitemShut {NoStop}%
\bibitem [{\citenamefont {Pikovsky}\ \emph {et~al.}(2003)\citenamefont
  {Pikovsky}, \citenamefont {Kurths}, \citenamefont {Rosenblum},\ and\
  \citenamefont {Kurths}}]{synchronisation}%
  \BibitemOpen
  \bibfield  {author} {\bibinfo {author} {\bibfnamefont {A.}~\bibnamefont
  {Pikovsky}}, \bibinfo {author} {\bibfnamefont {J.}~\bibnamefont {Kurths}},
  \bibinfo {author} {\bibfnamefont {M.}~\bibnamefont {Rosenblum}},\ and\
  \bibinfo {author} {\bibfnamefont {J.}~\bibnamefont {Kurths}},\ }\href
  {https://doi.org/10.1063/1.1554136} {\emph {\bibinfo {title}
  {Synchronization: a universal concept in nonlinear sciences}}},\ \bibinfo
  {number} {12}\ (\bibinfo  {publisher} {Cambridge university press},\ \bibinfo
  {year} {2003})\BibitemShut {NoStop}%
\bibitem [{\citenamefont {Stukowski}(2009)}]{stukowski2009visualization}%
  \BibitemOpen
  \bibfield  {author} {\bibinfo {author} {\bibfnamefont {A.}~\bibnamefont
  {Stukowski}},\ }\bibfield  {title} {\bibinfo {title} {Visualization and
  analysis of atomistic simulation data with ovito--the open visualization
  tool},\ }\href@noop {} {\bibfield  {journal} {\bibinfo  {journal} {Modelling
  and Simulation in Materials Science and Engineering}\ }\textbf {\bibinfo
  {volume} {18}},\ \bibinfo {pages} {015012} (\bibinfo {year}
  {2009})}\BibitemShut {NoStop}%
\bibitem [{\citenamefont {Schaller}\ and\ \citenamefont
  {Meyer-Hermann}(2005)}]{schaller2005multicellular}%
  \BibitemOpen
  \bibfield  {author} {\bibinfo {author} {\bibfnamefont {G.}~\bibnamefont
  {Schaller}}\ and\ \bibinfo {author} {\bibfnamefont {M.}~\bibnamefont
  {Meyer-Hermann}},\ }\bibfield  {title} {\bibinfo {title} {Multicellular tumor
  spheroid in an off-lattice voronoi-delaunay cell model},\ }\href
  {https://doi.org/10.1103/PhysRevE.71.051910} {\bibfield  {journal} {\bibinfo
  {journal} {Phys. Rev. E}\ }\textbf {\bibinfo {volume} {71}},\ \bibinfo
  {pages} {051910} (\bibinfo {year} {2005})}\BibitemShut {NoStop}%
\end{thebibliography}
%

\end{document}